\documentclass[fleqn,usenatbib]{mnras}
\usepackage{amsfonts}
\usepackage{amssymb}
\usepackage{newtxtext,newtxmath}
\usepackage[T1]{fontenc}
\DeclareRobustCommand{\VAN}[3]{#2}
\let\VANthebibliography\thebibliography
\def\thebibliography{\DeclareRobustCommand{\VAN}[3]{##3}\VANthebibliography}
\usepackage{graphicx}
\usepackage{caption}
\usepackage{comment}
\usepackage{subcaption}
\usepackage{amsmath}
\usepackage{mathtools}
\usepackage{color}
\definecolor{violet}{rgb}{0.5,0,0.5}
\definecolor{purple}{rgb}{0.5,0,0.5}
\definecolor{darkgreen}{rgb}{0.01,0.5,0.32}
\usepackage{bm}
\usepackage{braket}

\allowdisplaybreaks
\def\be{\begin{equation}}
\def\ee{\end{equation}}
\newcommand{\ud}{\mathrm{d}}
\newcommand{\ii}{\mathrm{i}}
\newcommand{\rme}{\mathrm{e}}

\title{Three-dimensional wave optics for weak-field lensing of gravitational waves}

\author[M. Carrillo Gonzalez et al.]{
Mariana Carrillo Gonzalez,$^{1,2}$\thanks{E-mail: m.carrillo-gonzalez@imperial.ac.uk}
Alan Heavens,$^{3}$
\thanks{E-mail: a.heavens@imperial.ac.uk}
Lorne Whiteway$^{4}$
\thanks{E-mail: lorne.whiteway.13@ucl.ac.uk}
\\
$^{1}$Abdus Salam Centre for Theoretical Physics, Imperial College, London, SW7 2AZ, U.K\\
$^{2}$School of Physics \& Astronomy, University of Southampton, Highfield, Southampton, SO17 1BJ, U.K\\
$^{3}$Imperial Centre for Inference and Cosmology (ICIC), Department of Physics, Imperial College, London SW7 2AZ, UK\\
$^{4}$Department of Physics and Astronomy, University College London, Gower Street, London WC1E 6BT, UK
}

\date{\today}
\pubyear{\the\year{}}

\begin{document}
\label{firstpage}
\pagerange{\pageref{firstpage}--\pageref{lastpage}}
\maketitle

% Abstract of the paper
\begin{abstract}
We develop a perturbative Green's function approach to gravitational lensing by weak gravitational potentials that need not be localized on a thin lens plane and that applies in both the wave optics and geometric optics regimes. We recast position-space integrals as Fourier, or momentum-space, integrals that appear in scattering amplitude calculations. The method gives the Born approximation directly in three dimensions and can be systematically extended to post-Born orders. For a Schwarzschild lens, we compute the leading Born term and new post-Born contributions arising from the order $\mathcal{O}(G^2)$ correction to the potential, keeping finite-distance corrections beyond the usual paraxial expansion. We show that these general-relativistic corrections are controlled by the parameter $GM\omega\,b/\chi_{\rm eff}$ in the small-angle regime, and are therefore negligible for standard weak-lensing configurations but become relevant in more extreme geometries (such as hierarchical triples with very small source--lens separations). We also discuss higher-order Newtonian corrections, their infrared sensitivity for a long-range potential, and the regulated form of the Newtonian potential given by the Yukawa potential. Finally, we formulate the corresponding calculation in an FLRW background, identifying the leading flat-space limit and estimating the size of curvature-induced corrections including tails. This method clarifies the regime of validity of the Born, large-distance, and paraxial approximations in gravitational-wave lensing and provides a framework for treating generic lensing potentials. 
\end{abstract}


%%%%%%%%%%%%%%%%%%%%%%%%%%%%%%%%%%%%%%%%%%%%%%%%%%

%%%%%%%%%%%%%%%%% BODY OF PAPER %%%%%%%%%%%%%%%%%%

\section{Introduction}
Our ability to observe gravitational waves has given us a new way of studying compact objects, their environments, and the intervening matter distribution between the source and the observer.  Gravitational lensing is an unavoidable part of this programme.  Just as with electromagnetic signals, gravitational waves are deflected, magnified, delayed, and sheared by gravitational potentials along the line of sight.  However, gravitational waves have long wavelengths and phase-coherent waveforms; this means that lensing can leave frequency-dependent amplitude and phase modulations that have no direct analogue in standard geometric-optics observations of light.  These effects are especially interesting for current and future detectors.  In the ground-based band, lensing can bias the inferred source-frame masses and redshifts of compact-binary mergers and can generate repeated or distorted signals.  In the space-based band, where the wavelengths are much larger, diffraction and interference can become part of the waveform itself.  Future observatories such as LISA, the Einstein Telescope, Cosmic Explorer, and DECIGO are therefore expected to be able to use the lensing of gravitational waves as a quantitative probe of compact objects, of dark matter substructure, and of cosmology, rather than simply a rare curiosity \citep{Takahashi:2003ix,Oguri:2018muv,Dai:2017huk,Li:2019rns,Urrutia:2021qak,Jung:2017flg}.

Most of the literature on gravitational-wave lensing uses formalisms with constructions in the lens plane.  In this approach the lens is projected onto a two-dimensional plane transverse to the line of sight, and the effect of the lens is encoded in a projected potential or time-delay function.  In the geometric-optics limit this gives the usual image positions, magnifications, and time delays, while in wave optics, the interference of various signals is considered, and the observed waveform is instead encoded in an amplification factor obtained from a two-dimensional diffraction integral over the lens plane. This framework, where the lens is projected to a plane, has been extremely useful and underlies most phenomenological studies of point lenses, galaxy lenses, microlensing, and multiple-lens systems \citep{Nakamura:1997sw,Takahashi:2003ix,Yamamoto:2003cd,Oguri:2018muv,Feldbrugge:2020tti,Ramesh:2021nnl}.  Nevertheless, the projection to a lens plane is an approximation.  It assumes a hierarchy in which the lens is thin compared with the source--observer distance and in which the relevant propagation problem can be effectively reduced to a two-dimensional one.  These assumptions are well motivated in many astrophysical applications, but they can break down in some regimes \citep{Suyama:2005mx,Frittelli:2011uh}.

Hierarchical triples provide a concrete setting where this limitation can become important. In such systems, the gravitational-wave source is a compact binary orbiting near a massive black hole, so the lens can be close to the source and the incident wavefront at the lens need not be well approximated as a plane wave. This has motivated studies of repeated lensing, strong-field propagation, spin-Hall corrections, and deep wave-optics effects in hierarchical triples \citep{DOrazio:2019fbq,Cardoso:2026ugm,Yu:2021dqx,Oancea:2022szu,Pijnenburg:2024btj}. Here we focus on a complementary observable: the finite-distance corrections to the wave-optics amplification factor, obtained by propagating the wave in three dimensions rather than imposing a lens-plane or asymptotic-scattering description.

There are a few approaches that go beyond, or test the limits of, the standard two-dimensional diffraction integral. One prominent case is the Born approximation \citep[introduced in][]{Takahashi:2005sxa} in which the lensed signal was derived from a three-dimensional Helmholtz equation by treating the gravitational potential perturbatively in the weak-field regime. This gives a genuinely three-dimensional integral representation before any projection is made.  As a first-order expansion in the gravitational potential, the Born approximation captures wave-optics effects that are linear in the lensing potential, but does not capture non-perturbative strong-lensing effects such as caustic diffraction, Einstein-ring-dominated amplification, or generic multi-image interference. For a compact lens, the computation is usually reduced to an integral over the projected two-dimensional potential, thereby reproducing the expansion at leading order in the lensing potential of the diffraction integral.  The same weak-field Born framework has also been applied to propagation through cosmological density perturbations, where the finite Fresnel scale suppresses the effect of inhomogeneities below the wave-optics resolution scale \citep{Takahashi:2005ug}, although there are some observables that can probe structures below the Fresnel scale \citep{Tanaka:2025nfw}. Recent work has revisited the Born and post-Born expansion in several settings, including single lens planes, point-mass shot noise, and stochastic density perturbations \citep{Mizuno:2022xxp,Yarimoto:2024uew,Tanaka:2025nfw}. While these analyses go beyond the leading thin-lens diffraction integral in different ways, their working formulations still retain at some stage either a projected lens-plane description or paraxial/small scattering angle approximations. 

A complementary direction expands around geometric optics. Post-geometric-optics methods describe finite-frequency corrections to null-geodesic propagation, including polarization distortions, helicity-dependent corrections, and birefringence effects in curved backgrounds \citep{Cusin:2019rmt,Dalang:2021qhu,Kubota:2023dlz,Oancea:2022szu}. These methods are well suited to wavelengths shorter than the curvature scale, but they are not designed to reproduce the full diffractive regime of the usual amplification factor.

A third class of recent work addresses compact lenses more directly using black-hole perturbation theory, scattering methods, or path-integral methods. These approaches can include effects that are absent from the scalar point-lens diffraction integral, such as horizon absorption, tensor polarization, helicity mixing, and strong-field scattering  \citep{Pijnenburg:2024btj,Braga:2024pik,Chan:2025wgz}. Related three-dimensional wave-scattering calculations have treated compact black-hole backgrounds using partial-wave methods \citep{Nambu:2015aea,Nambu:2019sqn,Motohashi:2021zyv,Willenborg:2023ixu,Li:2025lvl}. A remaining question, which is addressed in this paper, is how the weak-field limit of the standard two-dimensional diffraction integral is corrected by a fully three-dimensional wave-propagation calculation. In doing so, we also obtain a closed-form resummation of the finite-distance corrections to the Born integral. This resummation is distinct from the thin-lens Fresnel resummation of \cite{Takahashi:2005sxa}; it allows the inclusion of finite distance corrections to the Green's function prefactor non-perturbatively.

\begin{figure}
    \centering
    \includegraphics[width=\columnwidth]{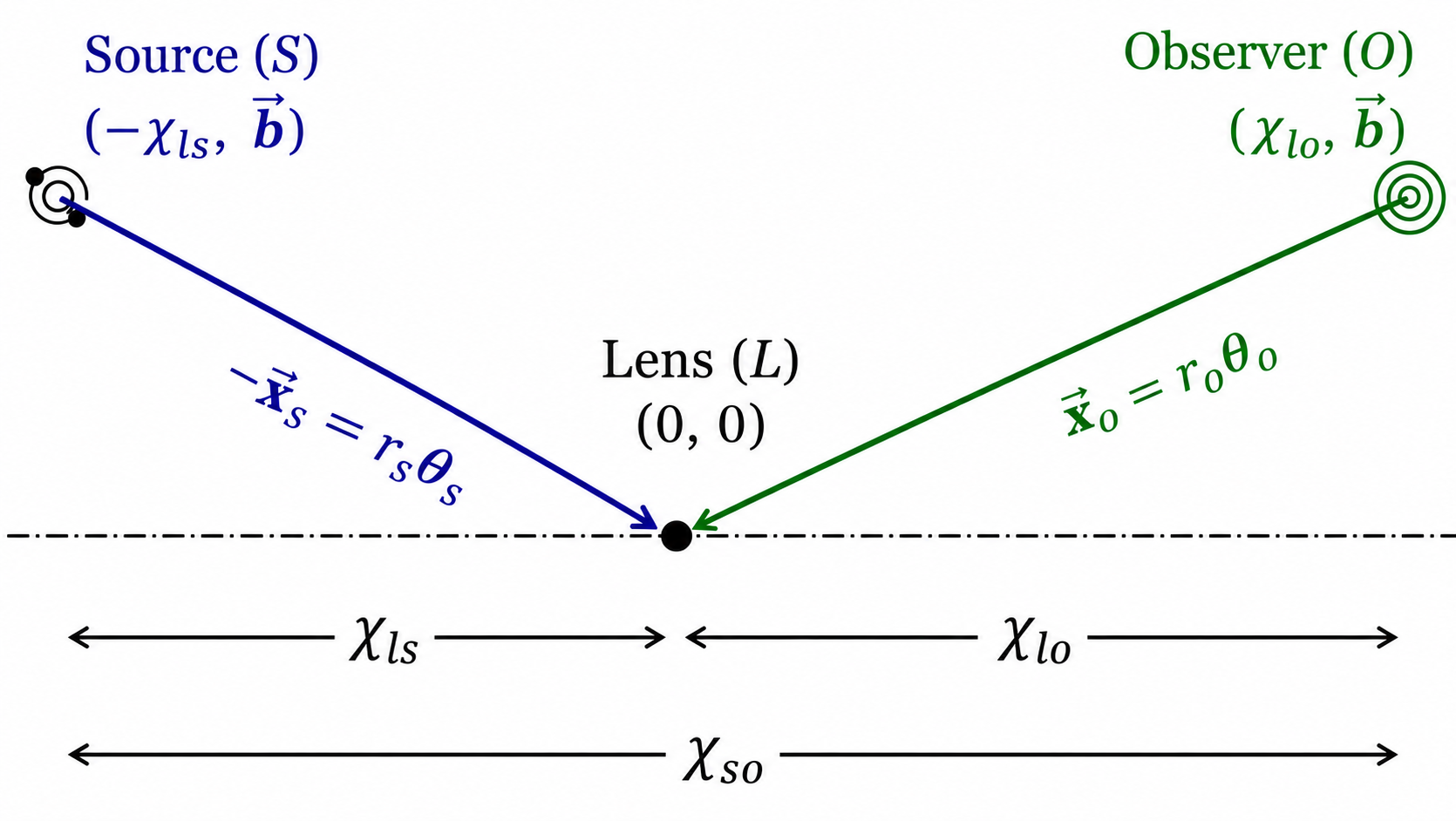}
    \caption{Geometrical setup for the three-dimensional lensing calculation. The lens is placed at the origin, with source and observer positions parametrized by the line-of-sight distances $\chi_{\rm ls}$ and $\chi_{\rm lo}$ and transverse separation $\mathbf b$.}
    \label{fig:geometrical-setup}
\end{figure}

For cosmological applications, the local lensing problem must be embedded in a Friedman-Lem\^{a}itre-Robertson-Walker (FLRW) spacetime, with the relevant geometry expressed in terms of angular-diameter distances and redshifted frequencies \citep{Ellis:1997ey,Schneider:1992book}. In the standard cosmological thin-lens treatment, these distance factors enter the lens equation, time-delay, and diffraction integral \citep{Schneider:1992book,Narayan:1996ba,Takahashi:2003ix}. From the point of view of three-dimensional wave propagation, the expansion of the background enters the wave operator and hence the retarded Green's function. These corrections (beyond the usual replacement of distances by appropriate angular diameter distances) are suppressed by powers of $H_0/\omega$ and for the frequencies of interest are negligible for most compact-lens applications.  Nevertheless, they are conceptually useful: they make clear which part of the waveform is modified by the local weak-field scattering and which part comes from propagation on the FLRW background.  The Green's function method used in this paper provides a systematic way of performing this separation and of adding such Hubble-scale corrections when needed. 

The paper is organized as follows.  In Section~\ref{sec:method} we introduce the Green's function method for wave propagation in a weak gravitational potential. In Section~\ref{sec:Schw} we specialize to a Schwarzschild lens, derive the effective potential through second post-Minkowskian order, and compute the Born and post-Born contributions to the waveform.  We separate the genuine post-Minkowskian correction to the potential from the Newtonian recursion term, since both enter at order $G^2$.  We then extend the setup to an expanding background and estimate the size of curvature and Hubble corrections. We analyze the regime of validity of the full three-dimensional calculation and its relation to the usual eikonal and paraxial approximations in Section~\ref{sec:observ}, as well as considering observational prospects. Finally, we conclude and discuss future directions in Section~\ref{sec:concl}. We leave the details of some calculations to the appendices (including an alternative derivation of our framework via path integrals, the calculation of the Fourier integrals used in the main text, and the finite-distance resummation).

%----------------------------
\section{Weak lensing: computational method} \label{sec:method}
We describe the computational method used to obtain the wave-optics amplification factor in the weak-field regime. We use the standard scalar-wave description of gravitational-wave lensing \citep{Schneider:1992book,Takahashi:2003ix}, in which the lens acts on the waveform through the curved-space propagation equation, and spin-dependent polarization effects are neglected. This treatment provides a controlled starting point for isolating the finite-distance and three-dimensional effects studied below. We therefore model the waveform by a scalar field $\phi$ satisfying $\nabla^2\phi=0$ on a background sourced by the gravitational potential $U$ of the lens, assumed to be weak, and we solve this equation perturbatively in $U$. It is useful to work at fixed frequency and decompose the scalar field into Fourier modes,
\begin{equation}
    \label{eq:fourier_phi}
    \phi(t,\mathbf{x})=\int \frac{d\omega}{2\pi}\,e^{-\ii \omega t}\phi_\omega(\mathbf{x}) \, ,
\end{equation}
so that each frequency mode propagates independently through the gravitational potential. Here and throughout the paper, boldface denotes three-dimensional vectors. The wave equation then reduces to the inhomogeneous Helmholtz equation (we take $c=1$ throughout)
\begin{equation}
(\boldsymbol{\partial}^2+\omega^2)\phi_{\omega}= 4 \omega^2 U \phi_{\omega} \ , \label{eq:inhom_Helmholtz_U}
\end{equation}
where $\boldsymbol{\partial}^2$ is the spatial three-dimensional Laplacian and $U$ is generically an operator, since it can involve derivatives (as seen below). 

We consider the setup shown in Fig.~\ref{fig:geometrical-setup}, where the lens is at the origin and positions are labeled by
\begin{equation}
\mathbf{r}=r\,\boldsymbol{\theta} \ ,
\end{equation}
where $\boldsymbol \theta$ is a unit vector.  Thus, the position of the observer and the position of the source are written respectively as
\begin{equation}
{\bf x}_{\rm O}=r_{\rm O} {\boldsymbol \theta_{\rm O}} \ , \quad {\bf x}_{\rm S}=-r_{\rm S} {\boldsymbol \theta_{\rm S}}\ ,
\end{equation}
where $\boldsymbol \theta_{\rm{O}}$ and $\boldsymbol \theta_{\rm{S}}$ are unit vectors. For a source located at $\mathbf{x}_{\rm S}$, we denote by $\phi_\omega(\mathbf{x},\mathbf{x}_{\rm S})$ the solution of Eq.~\eqref{eq:inhom_Helmholtz_U} evaluated at the point $\mathbf{x}$, with the second argument keeping track of the source boundary condition. The observed lensed waveform is then
\begin{equation}\label{eq:Greens}
\begin{split}
\phi_{\omega}&(\mathbf{x}_{\rm O},\mathbf{x}_{\rm S})
=\phi^{(0)}_{\omega}(\mathbf{x}_{\rm O},\mathbf{x}_{\rm S})  \\
&+4\omega^2
\int \ud{\bf x}\,
G_{\omega}^{(0)}({\bf x}_{\rm O},{\bf x})\,
U({\bf x})\,
\phi_{\omega}({\bf x},{\bf x}_{\rm S}) \ .
\end{split}
\end{equation}
Here $\phi^{(0)}_\omega$ is the unlensed waveform and $G_{\omega}^{(0)}({\bf x}_{\rm O},{\bf x})$ is the Green's function of the spacetime without the lens potential. This latter function satisfies
\be
\begin{split}
(\nabla^2+\omega^2)G_{\omega}^{(0)}({\bf x}_{\rm O},{\bf x})
={}&-\frac{\delta^3(\mathbf{x}_{\rm O}-\mathbf{x})}{\sqrt{-g}} \ ,
\end{split}
\label{eq:GreensEq}
\ee
where $g$ is the determinant of the metric in which the gravitational wave propagates, that is, the spacetime generated by the lens. Eq.~\eqref{eq:Greens} follows directly from the definition of the Green's function in Eq.~\eqref{eq:GreensEq} and can also be obtained from the path-integral construction of ~\cite{Braga:2024pik}, as shown in Appendix~\ref{app:BragaCalc}. In the next section, we keep the background static and suppress any explicit time dependence. We will explain in Section~\ref{sec:FLRW} how the same construction is modified when the propagation takes place on an FLRW background.

A useful observable is the complex amplification factor of the waveform,
\be
F=\frac{\phi_{\omega}(\mathbf{x}_{\rm O},\mathbf{x}_{\rm S})}
{\phi^{(0)}_{\omega}(\mathbf{x}_{\rm O},\mathbf{x}_{\rm S})} \ .
\ee
This quantity isolates the lens-induced change in the observed waveform: $F=1$ in the absence of the lens, while deviations from unity encode the phase shift and amplitude modulation generated by the potential.

\section{Static black hole lens} \label{sec:Schw}
We now specialize the general weak-lensing setup to a static Schwarzschild lens.  We start from the Schwarzschild metric in isotropic coordinates,
\begin{equation}
ds^2=-\left(\frac{1-\frac{GM}{2r}}{1+\frac{GM}{2r}}\right)^2 dt^2+\left(1+\frac{GM}{2r}\right)^4\left(dr^2+r^2d\Omega_2^2\right) \ ,
\end{equation}
where $d\Omega_2^2=d\theta^2+\sin^2\theta\,d\phi^2$, $G$ is Newton's gravitational constant, and $M$ is the mass of the lens. The scalar equation $\nabla^2\phi=0$, with $\nabla$ the covariant derivative associated with the Schwarzschild metric, can then be written as Eq.~\eqref{eq:inhom_Helmholtz_U} for each frequency mode $\phi_\omega$. Since the exact equation contains $GM/r$-dependent corrections multiplying the derivative terms, we solve it perturbatively in powers of $GM/r$. Concretely, at each order, we solve the equation for $\partial_r^2\phi_\omega$, and use the lower-order equation of motion to eliminate higher-order occurrences of $\partial_r^2\phi_\omega$. Iterating this procedure yields Eq.~\eqref{eq:inhom_Helmholtz_U} consistently at the desired order in $GM/r$. At second order in $GM/r$, the effective potential in Eq.~\eqref{eq:inhom_Helmholtz_U} contains the Newtonian term and General Relativity (GR) corrections,
\begin{equation} \label{eq:Uat2PMder}
4U=-\frac{4G M }{r}-\frac{15G^2 M^2}{2r^2}-\frac{G^2M^2}{2 \omega^2 r^3}\frac{\partial}{\partial r}+\mathcal{O}(G^3) \ .
\end{equation}
The simplest way to deal with the derivative term is to consider a field redefinition $\phi_{\omega}\rightarrow e^{\ii G^2 M^2/(8 r^2)}\phi_{\omega}$ which removes the derivative and gives instead the potential
\begin{equation} \label{eq:Uat2PM}
4U=-\frac{4G M }{r}-\frac{15G^2 M^2}{2r^2}-\frac{G^2M^2}{4 \omega^2 r^4}+\mathcal{O}(G^3) \ .
\end{equation}
We take the unlensed waveform to be an outgoing spherical wave emitted from the source,
\begin{equation}
\phi^{(0)}_\omega({\bf x}_{\rm O},{\bf x}_S) = C\,\frac{e^{\ii \omega |{\bf x}_{\rm O}-{\bf x}_S|}}{|{\bf x}_{\rm O}-{\bf x}_S|} \, . \label{eq:phi0}
\end{equation}
Here, $C$ is an overall normalization, which drops out of the amplification factor.

In the Newtonian approximation, we keep only the leading $\mathcal{O}(G)$ contribution to the potential and set $\phi=\phi^{(0)}$ on the right-hand side of Eq.~\eqref{eq:Greens}, which gives the Born approximation. To compute post-Born corrections, it is important to consider not only the recursion of the Newtonian potential, but also the higher-order in $G$ corrections to the potential $U$ arising from GR. Thus, the waveform including post-Born corrections at order $\mathcal{O}(G^2)$ (sometimes called 2nd post-Minkowskian) is given by
\begin{align}
\phi_{\omega}(\mathbf{x}_{\rm O},\mathbf{x}_{\rm S})
={}&
\phi^{(0)}_{\omega}(\mathbf{x}_{\rm O},\mathbf{x}_{\rm S})
\nonumber\\
&+4\omega^2
\int \ud{\bf x}'\,
G_{\omega}^{(0)}({\bf x}_{\rm O},{\bf x}')\,
U^{(1)}({\bf x}')\,
\phi^{(0)}_{\omega}({\bf x}',{\bf x}_{\rm S})
\nonumber\\
&+4\omega^2
\int \ud{\bf x}'\,
G_{\omega}^{(0)}({\bf x}_{\rm O},{\bf x}')\,
U^{(2)}({\bf x}')\,
\phi^{(0)}_{\omega}({\bf x}',{\bf x}_{\rm S})
\nonumber\\
&+16\omega^4
\int \ud{\bf x}'\,\ud{\bf x}''\,
G_{\omega}^{(0)}({\bf x}_{\rm O},{\bf x}')\,
G_{\omega}^{(0)}({\bf x}',{\bf x}'')
\nonumber\\
&\hspace{1.2cm}\times
U^{(1)}({\bf x}')U^{(1)}({\bf x}'')\,
\phi^{(0)}_{\omega}({\bf x}'',{\bf x}_{\rm S})\ ,
\label{eq:Greens_2PM}
\end{align}
where the flat space Green's function $G_{\omega}^{(0)}$ is
\begin{equation}\label{eq:FreeProp}
G^{(0)}_\omega ({\bf x},{\bf x}') = -\frac{1}{4 \pi} \frac{e^{\ii \omega |{\bf x}-{\bf x}'|}}{ |{\bf x}-{\bf x}'|} \ ,
\end{equation}
and $U^{(j)}$ is the $j$-th order in $G$ term of Eq.~\eqref{eq:Uat2PM}. For the computation of these integrals, we will assume the hierarchy of scales
\be \label{eq:approxs}
|\mathbf{x}'| \ll r_{\rm S}, r_{\rm O} \ ,
\ee
and clarify the meaning of this assumption.

\subsection{Born approximation} \label{sec:Born}
We start with the Born approximation, which consists of only the first two lines of
Eq.~\eqref{eq:Greens_2PM}. Writing
\begin{align}
\mathbf{k}=\omega(\boldsymbol{\theta}_{\rm S}-\boldsymbol{\theta}_{\rm O}) \ ,
\end{align}
and considering Eq.~\eqref{eq:approxs}, the integral becomes
\begin{align} \label{eq:born-integral}
\phi^{(1)}_{\omega}(\mathbf{x}_{\rm O}, \mathbf{x}_{\rm S})
&=
-\frac{\omega^2}{(4\pi)^2}\frac{4  G M}{r_{\rm O} r_{\rm S}}
e^{\ii\omega(r_{\rm O}+r_{\rm S})}\, I^{(1)} ,
\end{align}
with
\begin{align} \label{eq:FourierInt}
I^{(p)}
&\equiv
\int d^3\mathbf{r}\;
\frac{1}{|\mathbf{r}|^p}
\left(
1-\frac{\mathbf{r}\cdot\boldsymbol{\theta}_{\rm S}}{r_{\rm S}}
+\frac{\mathbf{r}\cdot\boldsymbol{\theta}_{\rm O}}{r_{\rm O}}
\right)
e^{\,\ii\mathbf{r}\cdot \mathbf{k}} \ .
\end{align}
We used Eq.~\eqref{eq:approxs} to expand the denominator and the phase. This integral is computed in Appendix~\ref{app:fourier} and in the present case gives
\begin{align}
I^{(1)}
&=
\frac{4\pi}{\mathbf{k}^2}
\left(
1-\frac{\ii}{\omega r_{\rm SO}} 
\right) =
\frac{4\pi}{2\omega^2\bigl(1-\boldsymbol{\theta}_{\rm S}\!\cdot\!\boldsymbol{\theta}_{\rm O}\bigr)}
\left(
1-\frac{\ii}{\omega r_{\rm SO}}
\right),
\end{align}
where we define
\begin{align}
\frac{1}{r_{\rm SO}} \equiv \frac{1}{r_{\rm S}}+\frac{1}{r_{\rm O}} \ .
\end{align}
Substituting back, we obtain
\begin{align}
\phi^{(1)}_{\omega}(\mathbf{x}_{\rm O}, \mathbf{x}_{\rm S})
&=
-\frac{ G M}{2\pi\, r_{\rm O} r_{\rm S} \bigl(1-\boldsymbol{\theta}_{\rm S}\!\cdot\!\boldsymbol{\theta}_{\rm O}\bigr)}
e^{\ii \omega(r_{\rm O}+r_{\rm S})}
\left(
1-\frac{\ii}{\omega r_{\rm SO}}
\right) \, ,
\end{align}
in agreement with \cite{Braga:2024pik}. This gives an amplification factor
\begin{equation} \label{eq:FBorn}
F=1+ \frac{ 2G M |r_{\rm O} {\boldsymbol \theta}_{\rm O}+r_{\rm S} {\boldsymbol \theta}_{\rm S}|}{r_{\rm O} r_{\rm S} \bigl(1-\boldsymbol{\theta}_{\rm S}\!\cdot\!\boldsymbol{\theta}_{\rm O}\bigr)} e^{\ii \Omega^{\rm rf}} \left( 1-\frac{\ii}{\omega r_{\rm SO}} \right) \, ,
\end{equation}
where we define the Fresnel phase
\begin{eqnarray}
\Omega^{\rm rf}\equiv \omega(r_{\rm O}+r_{\rm S}-|r_{\rm O} {\bf \theta}_{\rm O}+r_{\rm S} {\bf \theta}_{\rm S}|) \ .
\end{eqnarray}
It is now clear that by using Eq.~\eqref{eq:approxs} in the denominator of the integrand, we are considering an expansion in $(\omega r_{\rm SO})^{-1} \ll 1$. In Appendix~\ref{app:finite_distance_resummation}, we show how to resum these corrections. We give a closed-form resummation of the finite-distance (prefactor) corrections to the Born integral, which is distinct from the thin-lens Fresnel resummation, and which allows these effects to be included non-perturbatively:
\begin{equation} \label{eq:FBorn_resumFD}
F=1+ \frac{ 2G M |r_{\rm O} {\boldsymbol \theta}_{\rm O}+r_{\rm S} {\boldsymbol \theta}_{\rm S}|}{r_{\rm O} r_{\rm S} (1-\boldsymbol\theta_{\rm S}\cdot\boldsymbol\theta_{\rm O} )} e^{\ii \Omega^{\rm rf}} \int_0^\infty \!\!\ud\alpha \int_0^\infty \!\! \ud\beta\, \frac{ e^{-\alpha-\beta}}{ f(\alpha,\beta;\mathbf{x}_{\rm O},\mathbf{x}_{\rm S})} \, ,
\end{equation}
where
\begin{align}
f(\alpha,\beta;&\mathbf{x}_{\rm O},\mathbf{x}_{\rm S}) =  1 + \frac{\ii}{\omega}
\left( \frac{\alpha}{r_{\rm S}} + \frac{\beta}{r_{\rm O}} \right) \nonumber \\
&-\frac{1}{2\omega^2(1-\boldsymbol\theta_{\rm S}\cdot\boldsymbol\theta_{\rm O} )}\left( \frac{\alpha^2}{r_{\rm S}^2} + \frac{\beta^2}{r_{\rm O}^2} - 2\boldsymbol\theta_{\rm S}\cdot\boldsymbol\theta_{\rm O} \frac{\alpha\beta}{r_{\rm S} r_{\rm O}}
\right) \ .
\end{align}
We note that these finite distance effects are not captured by the diffraction integral, since they are neglected in the \textit{eikonal approximation}.

To make contact with the standard literature, we can rewrite our three dimensional vectors in terms of two dimensional quantities living in the lens plane. We align the source-observer line of sight with the $z$ axis such that
\begin{equation} \label{eq:3dto2d}
{\bf{x}}_{\rm O}=(\chi_{\rm lo},\Vec{b}) \ , \quad {\bf{x}}_{\rm S}=(-\chi_{\rm ls},\Vec{b}) \ .
\end{equation}
where we use an arrow to denote two-dimensional vectors. Assuming that the finite distance corrections are small and that the line-of-sight distances are larger than the transverse ones, that is, $b=|\Vec b|\ll \chi_{\rm lo},\chi_{\rm ls}$ turns the amplification factor into
\begin{equation}
F= 1+\frac{r_{\rm E}^2}{|\Vec b|^2} \exp\!\left(\ii\frac{|\Vec b|^2}{2r_{\rm F}^2}\right) \left( 1-\ii\frac{r_{\rm F}^2}{\chi_{\rm eff}^2} \right) \ , \label{eq:BornSchwBsmall}
\end{equation}
where we define the effective distance in the lens plane as
\be
\chi_{\rm eff}=\frac{\chi_{\rm lo}\chi_{\rm ls}}{\chi_{\rm so}} \ ,
\ee
with $\chi_{\rm so}=\chi_{\rm ls}+\chi_{\rm lo}$, and we introduced the Fresnel scale and Einstein radius, which are given as
\begin{equation}
r_{\rm F}=\sqrt{\frac{\chi_{\rm eff}}{\omega}}, \qquad r_{\rm E}=\sqrt{4GM\chi_{\rm eff}} \ .
\end{equation}
This small-angle limit is usually referred to as the \textit{paraxial approximation}. We can see that Eq.~\eqref{eq:BornSchwBsmall} indeed agrees with the result of \cite{Takahashi:2005sxa} in the limit $r_{\rm F} \lesssim |\Vec b| \ll \chi_{\rm eff}$, and after projecting the distances to the source plane.

\subsection{Fresnel phase corrections} \label{sec:Fresnel}
In deriving Eq.~\eqref{eq:born-integral}, we kept only the term linear in the integration point in the propagation phase. The expansion up to next leading order gives
\begin{equation} \label{eq:phase_exp}
\omega\left(|{\bf x}_{\rm O}-{\bf r}|+|{\bf r}-{\bf x}_{\rm S}|\right) = \omega(r_{\rm O}+r_{\rm S})+{\bf k}\cdot{\bf r} +\frac{\omega}{2}r_iQ_{ij}r_j+\cdots ,
\end{equation}
with
\begin{equation}
Q_{ij}=\frac{\delta_{ij}-\theta_{{\rm O} i}\theta_{\rm{O} j}}{ r_{\rm O}} +\frac{\delta_{ij}-\theta_{\rm{S} i}\theta_{\rm{S} j}}{ r_{\rm S}}.
\end{equation}
Thus, we have implicitly assumed that $\omega r_iQ_{ij}r_j$ is small. We proceed with this assumption and write the corrections to $I_1^{(1)}$ as 
\begin{equation} \label{eq:Fresnelphase_series}
\widehat{I}^{(1)} \simeq \frac{2\pi}{\omega^2(1-\boldsymbol\theta_{\rm S}\cdot\boldsymbol\theta_{\rm O})} \left( 1+\Delta_2+\Delta_3+\Delta_{22}+\Delta_4+\cdots \right) \ ,
\end{equation}
where the $\Delta_n$ terms come from the expansion of the phase at order $r^n$ and $\Delta_{22}$ from the quadratic term in the series expansion of the exponential with phase of order $r^2$. The leading correction can be straightforwardly computed and gives
\begin{align}
 \frac{2\pi\Delta_2}{\omega^2(1-\boldsymbol\theta_{\rm S}\cdot\boldsymbol\theta_{\rm O})}
 &=
 \frac{\ii \omega}{2}Q_{ij}
 \int d^3{\bf r}\,\frac{r_ir_j}{r}e^{\ii{\bf k}\cdot{\bf r}} \nonumber\\
 &= -\frac{\ii \omega}{2}Q_{ij}
 \frac{\partial^2}{\partial k_i\partial k_j}
 \left(\frac{4\pi}{k^2}\right) \ ,
\end{align}
and the higher-order corrections are found in Appendix~\ref{app:fresnel}. Here, we ignore the $(\omega r_{\rm SO})^{-1}$ corrections. After some algebra, we find
\begin{equation}
\Delta_2 = -\frac{\ii}{\omega r_{\rm SO}} \frac{\boldsymbol\theta_{\rm S}\cdot\boldsymbol\theta_{\rm O}}{ 1-\boldsymbol\theta_{\rm S}\cdot\boldsymbol\theta_{\rm O}} \simeq -2 \ii \,\frac{r_{\rm F}^2}{b^2} \ ,
\end{equation}
where the last approximation uses the small-angle limit. Thus, in the paraxial approximation, Eq.~\eqref{eq:BornSchwBsmall} is the leading term in a $r_{\rm F}^2/b^2$ expansion, as known from \cite{Takahashi:2005sxa}. 

The full three dimensional amplification factor, including the leading finite distance and Fresnel phase corrections, is 
\begin{equation}
F=1+ \frac{ 2G M |r_{\rm O} {\boldsymbol \theta}_{\rm O}+r_{\rm S} {\boldsymbol \theta}_{\rm S}|}{r_{\rm O} r_{\rm S} \bigl(1-\boldsymbol{\theta}_{\rm S}\!\cdot\!\boldsymbol{\theta}_{\rm O}\bigr)} e^{\ii\Omega^{\rm rf}} \left[ 1-\frac{\ii}{\omega r_{\rm SO}}\left(1+ \frac{\boldsymbol\theta_{\rm S}\cdot\boldsymbol\theta_{\rm O}}{1-\boldsymbol\theta_{\rm S}\cdot\boldsymbol\theta_{\rm O}}\right) \right] \, .
\end{equation}
which in the small-angle limit is
\begin{equation}
F= 1+\frac{r_{\rm E}^2}{|\Vec b|^2} \exp\!\left(\ii\frac{|\Vec b|^2}{2r_{\rm F}^2}\right) \left( 1-\ii\frac{r_{\rm F}^2}{\chi_{\rm eff}^2}-2\ii\frac{r_{\rm F}^2}{|\Vec b|^2} \right) \ . \label{eq:BornSchwBsmall_rfcorrec}
\end{equation}
We remark again that the full three dimensional results cannot be obtained from the diffraction integral, so these new Fresnel phase corrections can only be compared to known results when taking the small-angle limit.

We have seen that using the approximation in Eq.~\eqref{eq:approxs} for the denominators of the Green's functions gives an expansion in $(\omega r_{\rm SO})^{-1} \ll 1$, while using it in the phase, some of the corrections can get enhanced by a factor $\boldsymbol\theta_{\rm S}\cdot\boldsymbol\theta_{\rm O}/(1-\boldsymbol\theta_{\rm S}\cdot\boldsymbol\theta_{\rm O})$ in the paraxial limit. In Appendix~\ref{app:fresnel}, we show how to compute subleading corrections in Eq.~\eqref{eq:Fresnelphase_series} up to quartic order and how to resum them in the small-angle limit to obtain the result from \cite{Takahashi:2005sxa} which arises from expanding the diffraction integral 
%\citep{Takahashi:2005sxa} 
at leading order in the lens potential.

\subsection{Post-Born GR corrections}
The third line of Eq.~\eqref{eq:Greens_2PM}, which corresponds to the first GR correction, is computed analogously to the Born approximation. In this case, we have
\begin{align}
\phi^{(2)}_{\omega}(\mathbf{x}_{\rm O}, \mathbf{x}_{\rm S})
&=
-\frac{\omega^2}{(4\pi)^2}\frac{G^2 M^2}{2r_{\rm O} r_{\rm S}}
e^{\ii\omega(r_{\rm O}+r_{\rm S})}\, \left( 15 I^{(2)}+\frac{1}{4\omega^2}I^{(4)}\right) .
\end{align}
From Appendix~\ref{app:fourier} we obtain
\begin{align}
I^{(2)} &=
\frac{2\pi^2}{\omega \sqrt{2\bigl(1-\boldsymbol{\theta}_{\rm S}\!\cdot\!\boldsymbol{\theta}_{\rm O}\bigr)}}
\left(1-\frac{\ii}{2\omega r_{\rm SO}}\right) \, , \\
I^{(4)}
&= -\pi^2 \omega \sqrt{2\bigl(1-\boldsymbol{\theta}_{\rm S}\!\cdot\!\boldsymbol{\theta}_{\rm O}\bigr)}
\left(
1+\frac{\ii}{2\omega r_{\rm SO}}
\right) \, .
\end{align}
Therefore
\be
\begin{split}
\phi^{(2)}_{\omega}&(\mathbf{x}_{\rm O},\mathbf{x}_{\rm S})
=
-\frac{15 G^2 M^2\,\omega}{16\,r_{\rm O} r_{\rm S}\,
\sqrt{2\bigl(1-\boldsymbol{\theta}_{\rm S}\!\cdot\!\boldsymbol{\theta}_{\rm O}\bigr)}}
\,e^{\ii\omega(r_{\rm O}+r_{\rm S})}
\\
&\times
\left[
\left(1-\frac{\ii}{2\omega r_{\rm SO}}\right)
-\frac{1-\boldsymbol{\theta}_{\rm S}\cdot\boldsymbol{\theta}_{\rm O}}{60}
\left(1+\frac{\ii}{2\omega r_{\rm SO}}\right)
\right].
\end{split}
\ee
which gives a correction to the amplification factor of the form 
\be \label{eq:F_PB_GR}
\begin{split}
\Delta F^{\rm GR}&
=
\frac{15 \pi G^2 M^2\,\omega \, \chi_{\rm so}}{4\,r_{\rm O} r_{\rm S}\,
\sqrt{2\bigl(1-\boldsymbol{\theta}_{\rm S}\!\cdot\!\boldsymbol{\theta}_{\rm O}\bigr)}}
\,e^{\ii\Omega^{\rm rf}}
\\
&\times
\left[
\left(1-\frac{\ii}{2\omega r_{\rm SO}}\right)
-\frac{1-\boldsymbol{\theta}_{\rm S}\cdot\boldsymbol{\theta}_{\rm O}}{60}
\left(1+\frac{\ii}{2\omega r_{\rm SO}}\right)
\right] \ .
\end{split}
\ee
If desired, the finite distance correction can be resummed as shown in Appendix~\ref{app:finite_distance_resummation}.  In the small-angle limit, where the result reduces to the standard two-dimensional lens-plane description, we recover the GR corrections analyzed in \cite{CarrilloGonzalez:2025gqm}. This new result follows from simple three-dimensional Fourier transforms of the effective wave equation, rather than from evaluating one-loop scattering integrals.

Having computed these leading-order, full three-dimensional, GR corrections, we can proceed to understand the scale at which higher-order GR corrections arise. The GR corrections to the potential are of two forms:
\be
U\sim\left(\frac{GM}{r}\right)^p \, ,  \,  \qquad U\sim\frac{1}{\omega^2 r^2}\left(\frac{GM}{r}\right)^p \, , \quad  \ \text{for} \  p \in \mathbb{Z}, \ p\geq2
\ee
where the one on the RHS arises from derivatives in the equation of motion. When $p$ is even, $p=2m$, $m\geq1$, we have that these terms contribute to the waveform as
\begin{align}
\phi_\omega^{\rm GR}
&\sim
\frac{(GM\omega)^{2m}}
{\omega r_{\rm SO}(r_{\rm O}+r_{\rm S})}
\left(1-\boldsymbol{\theta}_{\rm S}\!\cdot\!\boldsymbol{\theta}_{\rm O}\right)^{m-\frac{3}{2}},
\nonumber\\
\phi_\omega^{{\rm GR}\,{\rm deriv.}}
&\sim
\frac{(GM\omega)^{2m}}
{\omega r_{\rm SO}(r_{\rm O}+r_{\rm S})}
\left(1-\boldsymbol{\theta}_{\rm S}\!\cdot\!\boldsymbol{\theta}_{\rm O}\right)^{m-\frac{1}{2}},
\end{align}
where we have ignored additional terms suppressed by inverse powers of $\omega r_{\rm SO}$.  From this, we can see that the suppression of Post-Born GR terms with respect to the leading Born approximation is
\begin{align}
\frac{\phi_\omega^{\rm GR}}{\phi_\omega^{(1)}}
&\sim
(GM\omega)^{2m-1}
\left(1-\boldsymbol{\theta}_{\rm S}\!\cdot\!\boldsymbol{\theta}_{\rm O}\right)^{m-\frac{1}{2}},
\nonumber\\
\frac{\phi_\omega^{{\rm GR}\,{\rm deriv.}}}{\phi_\omega^{(1)}}
&\sim
(GM\omega)^{2m-1}
\left(1-\boldsymbol{\theta}_{\rm S}\!\cdot\!\boldsymbol{\theta}_{\rm O}\right)^{m+\frac{1}{2}}.
\end{align}

In the paraxial approximation, where line-of-sight distances are much larger than the transverse ones, the GR Post-Born corrections scale as
\begin{align}
\frac{\phi_\omega^{\rm GR}}{\phi_\omega^{(1)}}
&\sim
\left(GM\omega\frac{|\Vec{b}|}{\chi_{\rm eff}}\right)^{2m-1},
\nonumber\\
\frac{\phi_\omega^{{\rm GR}\,{\rm deriv.}}}{\phi_\omega^{(1)}}
&\sim
\left(GM\omega\frac{|\Vec{b}|}{\chi_{\rm eff}}\right)^{2m-1}
\left(\frac{|\Vec{b}|}{\chi_{\rm eff}}\right)^2 .
\end{align}
The correction that arises from what would be first derivatives of the field in the original equation of motion is suppressed by additional powers of $|\Vec{b}|/\chi_{\rm eff}$. Thus, GR corrections are suppressed as long as
\begin{equation}
\epsilon^{\rm GR}\equiv GM\omega \sqrt {\left(1-\boldsymbol{\theta}_{\rm S}\!\cdot\!\boldsymbol{\theta}_{\rm O}\right)} \sim GM\omega \frac{|\Vec{b}|}{\chi_{\rm eff}} \ll 1 .
\end{equation}
This means that they become relevant and the Born approximation breaks down only if $GM\omega$ is very large, i.e., deep in the geometric-optics regime, or sooner if the source, lens, and observer are not closely aligned, that is, if $\boldsymbol{\theta}_{\rm S}\!\cdot\!\boldsymbol{\theta}_{\rm O}\ll 1$.

\subsection{Post-Born Newtonian recursion corrections}
\label{sec:Newt}
For the recursion of the Newtonian term in the last line of Eq.~\eqref{eq:Greens_2PM}, the direct Born iteration develops an infrared divergence at large radius. This is not a local breakdown of the weak-field expansion, but rather a consequence of the long-range nature of the Coulomb potential. The usual definition of the amplification factor assumes a short-range lensing potential, so that far from the lens one may separate the solution into the unlensed spherical wave plus a scattered contribution with the same free asymptotics:
\be
\phi_{\omega}(\mathbf{x}_{\rm O},\mathbf{x}_{\rm S})
=
\frac{e^{\ii\omega|\mathbf{x}_{\rm O}-\mathbf{x}_{\rm S}|}}
{|\mathbf{x}_{\rm O}-\mathbf{x}_{\rm S}|}
+
(F_{\rm diff}-1)
\frac{e^{\ii\omega|\mathbf{x}_{\rm O}-\mathbf{x}_{\rm S}|}}
{|\mathbf{x}_{\rm O}-\mathbf{x}_{\rm S}|}\ .
\ee
This asymptotic form is not appropriate for long-range potentials such as the Newtonian one, for which the wave acquires the usual Coulomb phase at large radius. This is a well-known issue in Coulomb scattering \citep{Schwinger:1964zzb,KADYROV20091516}, and several prescriptions exist for treating the corresponding infrared divergences in scattering calculations \citep{Dollard:1964cok,Kulish:1970ut,Forde:2003jt,Hannesdottir:2019opa,Hannesdottir:2019umk,Lippstreu:2023vvg}.

In the present case, this difficulty can be avoided because the point-mass Newtonian potential is exactly solvable \citep{Peters:1974gj,Takahashi:2003ix,Hostler:1963zz}. Defining
\be
X=r_{\rm O}+r_{\rm S}+|\mathbf{x}_{\rm O}-\mathbf{x}_{\rm S}|,\qquad Y=r_{\rm O}+r_{\rm S}-|\mathbf{x}_{\rm O}-\mathbf{x}_{\rm S}|,
\ee
the Newtonian amplification factor can be written compactly as
\begin{align}
F_{\rm Newt.}&=
-\frac{e^{- \ii \omega |\mathbf{x}_{\rm O}-\mathbf{x}_{\rm S}|}}{\ii \omega}\,
\Gamma(1-2 \ii GM\omega)
\nonumber\\
&\mathrel{\phantom{=}}{}\times
(\partial_X-\partial_Y)
\left[
W_{2 \ii GM\omega,\frac12}(- \ii \omega X)
M_{2 \ii GM\omega,\frac12}(- \ii \omega Y)
\right] \ ,
\label{eq:exactNewt}
\end{align}
where $W_{\kappa,\mu}(z)$ and $M_{\kappa,\mu}(z)$ are the Whittaker functions. This is the Hostler–Pratt closed form of the Coulomb Green's function \citep{Hostler:1963zz}. To the best of our knowledge it has not previously been used to express the finite-distance wave-optics amplification factor for gravitational waves, where the usual treatment projects onto a thin lens.

Thus, the computation of the Post-Born Newtonian term is not very illuminating, since it can be obtained by expanding the known result. By expanding the exact small-angle solution in \cite{Takahashi:2003ix}, one can see that there are two contributions: $GM\omega (r_{\rm E}/b)^2 $ and $(GM\omega )^2$. The former was estimated in \cite{Yarimoto:2024uew}, and
dominates if $b\lesssim r_{\rm F}$. Since the post-Born corrections have to be smaller than the Born approximation, consistency requires
\begin{equation} \label{eq:Newt_scales}
GM\omega\,\frac{r_{\rm E}^2}{b^2} \ , \ (GM\omega)^2 \ll \frac{r_{\rm E}^2}{b^2}\ll 1.
\end{equation}
If we want to access regimes in which $GM\omega$ is not perturbatively small, we should resum the Newtonian interaction by using the exact solution in Eq.~\eqref{eq:exactNewt}. The genuinely relativistic  corrections can then be added perturbatively.

We now replace the Newtonian potential by a Yukawa potential. The finite range of the Yukawa interaction regulates the large-distance behavior of the Born integrals, while leaving a genuinely nontrivial post-Born lensing correction. This gives a new analytically tractable example of finite-distance wave-optics lensing and provides a simple setting in which to demonstrate the power of the method: the post-Born correction can be computed directly using scattering-amplitude loop-integration techniques, without relying on the usual two-dimensional lens-plane reduction. Explicitly, we take the leading potential to be the Yukawa deformation
\be
U^{(1)}=-\frac{G M}{r}e^{-r\kappa} \, ,
\ee
where $\kappa$ is the inverse Yukawa range, and work in the regime $\omega/\kappa\gg1$. The contribution obtained by inserting this potential twice in the Born series can be written as
\begin{align}
(\phi^{(2)}_{\omega})^{(U^{(1)})^2}(\mathbf{x}_{\rm O}, \mathbf{x}_{\rm S}) &=
-\omega^2\frac{( G M \omega)^2}{4\pi^3 r_{\rm S} r_{\rm O}} e^{\ii\omega(r_{\rm O}+r_{\rm S})} I^\text{Yuk.}  \,
\label{eq:G2PMfromU1sq}
\end{align}
with
\begin{align}
I^{\rm Yuk.}
\simeq{}&
\int \ud^3r'\,\ud^3r''\,
\frac{\rme^{-\kappa r'}}{r'}\,
\frac{\rme^{-\kappa r''}}{r''}\,
\frac{\rme^{\ii\omega|{\bf r}'-{\bf r}''|}}
{|{\bf r}'-{\bf r}''|}
\nonumber\\
&\times
\rme^{\ii{\bf q}_{\rm O}\cdot{\bf r}'}
\rme^{\ii{\bf q}_{\rm S}\cdot{\bf r}''}
\left(
1-\frac{\boldsymbol\theta_{\rm O}\!\cdot{\bf r}'}{r_{\rm O}}
-\frac{\boldsymbol\theta_{\rm S}\!\cdot{\bf r}''}{r_{\rm S}}
\right).
\end{align}
The useful point of this example is that the position-space convolution can be recast as a standard momentum-space loop integral. Using the momentum-space representation for the outgoing Green's function
\begin{align}
\frac{e^{\ii\omega|{\bf x}|}}{|{\bf x}|}
=\frac{4\pi}{(2\pi)^3}\int d^3k\;
\frac{e^{\ii{\bf k}\cdot{\bf x}}}{k^2-(\omega+\ii0)^2},
\label{eq:helmholtz-fourier}
\end{align}
together with the Yukawa transform
\begin{align}
\int d^3r\;\frac{e^{-\kappa r}}{r}\,e^{\ii{\bf p}\cdot{\bf r}}
=\frac{4\pi}{p^2+\kappa^2}\ ,
\label{eq:yukawa-ft}
\end{align}
the leading part of $I^{\rm Yuk.}$ is controlled by the scalar master integral
\begin{equation}
\begin{split}
I_0(q_{\rm O},q_{\rm S};\omega,\kappa)
={}&
8\int\ud^3 k\,
\frac{1}{k^2-(\omega+\ii0)^2}
\\
&\times
\frac{1}{(k+q_{\rm O})^2+\kappa^2}\,
\frac{1}{(k-q_{\rm S})^2+\kappa^2}.
\end{split}
\end{equation}
The terms linear in ${\bf r}'/r_{\rm O}$ and ${\bf r}''/r_{\rm S}$ are then generated by differentiating the same master integral with respect to the external angular variables. Thus
\begin{align}
I^{\rm Yuk.}
= I_0 + \frac{\ii}{\omega r_{\rm S}}\boldsymbol\theta_{\rm S}\!\cdot\!\frac{\partial I_0}{\partial \boldsymbol\theta_{\rm S}}+ \frac{\ii}{\omega r_{\rm O}}\boldsymbol\theta_{\rm O}\!\cdot\!\frac{\partial I_0}{\partial \boldsymbol\theta_{\rm O}}.
\end{align}
This form makes explicit why amplitude loop-integration techniques are useful: the finite-distance corrections are obtained from derivatives of a single three-propagator master integral rather than from a new position-space calculation.

One can solve for $I_0$ following \cite{Davydychev:1997wa} and then take the large $\omega/\kappa$ limit to find
\be
I_0
=
\frac{4 \ii \pi^2}{\omega^3\bigl(1+\bm\theta_{\rm O}\!\cdot\!\bm\theta_{\rm S}\bigr)}
\ln\!\left[\frac{2\omega^2}{\kappa^2}\left(1+\bm\theta_{\rm O}\!\cdot\!\bm\theta_{\rm S}\right)\right] \, .
\ee
Note that we no longer have a divergence in the forward limit $\bm\theta_{\rm O}\cdot\bm\theta_{\rm S}=1$, which appears when dealing with long-range forces, since we are considering the Yukawa potential. With this we find
\begin{equation}
\begin{split}
(\phi^{(2)}_{\omega})^{(U^{(1)})^2}
=&
-\frac{GM}{\pi r_{\rm O}r_{\rm S}}\,
\rme^{\ii(r_{\rm O}+r_{\rm S})\omega}
\left(
\ii GM\omega\,
\frac{
\ln\!\left[\frac{2\omega^2}{\kappa^2}
(1+\bm\theta_{\rm O}\!\cdot\!\bm\theta_{\rm S})\right]
}{
1+\bm\theta_{\rm O}\!\cdot\!\bm\theta_{\rm S}
}\right.
\\
&
\left. -\frac{GM}{r_{\rm SO}}\,
\frac{
(\bm\theta_{\rm O}\!\cdot\!\bm\theta_{\rm S})
\left\{
1-\ln\!\left[\frac{2\omega^2}{\kappa^2}
(1+\bm\theta_{\rm O}\!\cdot\!\bm\theta_{\rm S})\right]
\right\}
}{
(1+\bm\theta_{\rm O}\!\cdot\!\bm\theta_{\rm S})^2
}
\right) \ .
\end{split}
\end{equation}
Physically, this infrared-regulated example shows how the leading correction is controlled by the phase accumulated through two successive scatterings off the lens, and illustrates that the finite-distance post-Born terms can be organized efficiently in terms of standard momentum-space master integrals and their derivatives.

%----------------------------
\section{Weak lensing in FLRW}
\label{sec:FLRW}

\subsection{Green's function in the weak gravitational coupling regime of an FLRW universe}

We now extend the weak-field construction to propagation on an expanding FLRW background. For the frequencies of interest, corrections controlled by the ratio $H/\omega$ are expected to be negligible. Nevertheless, the present formalism allows this expectation to be checked directly. We do this through a systematic treatment of wave-optics lensing based on the Hadamard form of the Green's function, which makes the local propagation structure and the cosmological distance dependence explicit. This gives a first-principles derivation, away from the $H/\omega\rightarrow0$ limit, of the usual cosmological prescription where flat-space distances are replaced by the appropriate angular-diameter distances \citep{Takahashi:2005ug}.

We work in conformal coordinates and take the lens to be described by a weak, time-independent potential, so that
\be
ds^2=-a^2(\eta)\{[1+2U(\mathbf{x})]d\eta^2+[1-2U(\mathbf{x})]d\mathbf{x}^2\} .
\ee
The case of geometric optics in an expanding universe has been analyzed in \cite{Piattella:2015xga,Bessa:2022sdh,Takizawa:2021jxa}. Here, instead, we apply the wave-optics method of Section~\ref{sec:method}. The only new ingredient is the appropriate Green's function on the FLRW background.

Given the conformal structure of the metric, it is useful to factor out the scale factor by writing
\be
\phi_\omega(\mathbf{x},\eta)=\frac{1}{a(\eta)}\varphi(\mathbf{x},\eta). \label{eq:phi_frw}
\ee
At leading order in $U$, this field satisfies 
\begin{equation}\label{eq:eomFLRW}
\left[\boldsymbol{\partial}^2+(1-4U)\left(-\partial_\eta^2+\frac{a^2R^{\rm FLRW}}{6}\right)\right]\varphi(\mathbf{x},\eta)=0,
\end{equation}
where, as before, $\boldsymbol{\partial}^2$ is the spatial Laplacian, and $R^{\rm FLRW}$ is the Ricci scalar of the FLRW spacetime. The corresponding FLRW Green's function can be written as
\be
G(\mathbf{x}_{\rm O},\eta_{\rm O},\mathbf{x}_{\rm S},\eta_{\rm S})=\frac{1}{a(\eta_{\rm O})a(\eta_{\rm S})}g(\mathbf{x}_{\rm O},\eta_{\rm O},\mathbf{x}_{\rm S},\eta_{\rm S}),
\ee
where $g$ satisfies
\begin{align}
\mathcal{L}_{\rm FLRW} \ g(\mathbf{x}_{\rm O},\eta_{\rm O},\mathbf{x}_{\rm S},\eta_{\rm S})
=-\frac{\delta(\eta_{\rm O}-\eta_{\rm S})\delta^3(\mathbf{x}_{\rm O}-\mathbf{x}_{\rm S})}{\sqrt{-g}},
\end{align}
with
\be
\mathcal{L}_{\rm FLRW}\equiv-\partial_\eta^2+\boldsymbol{\partial}^2+\frac{a^2R^{\rm FLRW}}{6}.
\ee
Thus, the expansion of the Universe is isolated in the scale-factor prefactors and in the curvature term of the reduced Green's function equation, while the lens potential continues to enter perturbatively as in the static calculation.

The reduced Green's function has the Hadamard form
\begin{align}
&g(\mathbf{x}_{\rm O},\eta_{\rm O},\mathbf{x}_{\rm S},\eta_{\rm S})
= \nonumber \\
&\frac{1}{4\pi}\Bigg[ \frac{\delta(-\Delta\eta+\Delta x)}{\Delta x} +B(\eta_{\rm O},\eta_{\rm S})\Theta(-\Delta\eta+\Delta x) \Bigg] \ ,
\end{align}
where $\Delta\eta=\eta_{\rm O}-\eta_{\rm S}$ and $\Delta x=|\mathbf{x}_{\rm O}-\mathbf{x}_{\rm S}|$. The first term is the usual light-cone contribution. The second term is the tail contribution: unlike in flat space, propagation on an FLRW background has support inside the light cone because curvature scatters the wave. The tail function $B$ satisfies \citep{Burko:2002ge,Haas:2004kw}
\be
\left(-\partial_\eta^2+\boldsymbol{\partial}^2+\frac{a^2R^{\rm FLRW}}{6}\right)B=0,
\ee
\be
\left(\Delta x^\mu\partial_\mu B+B-\frac{a^2R^{\rm FLRW}}{12}\right)\bigg|_{\Delta\eta=\Delta x}=0,
\ee
together with the coincidence condition
\be
\lim_{\eta_1 \rightarrow \eta_2}B(\eta_1,\eta_2)=\frac{1}{12}a^2R^{\rm FLRW}.
\ee

We now transform to frequency space by Fourier transforming in $\Delta\eta$. This separates the Green's function into the flat-space-like light-cone piece and the curvature-induced tail,
\begin{equation}\label{eq:GreenFRW}
g^{(0)}_\omega(\mathbf{x}_{\rm O},\mathbf{x}_{\rm S};\eta)=g^{(0)}_{\rm lc}(\mathbf{x}_{\rm O},\mathbf{x}_{\rm S};\eta)+g^{(0)}_{\rm tail}(\mathbf{x}_{\rm O},\mathbf{x}_{\rm S};\eta),
\end{equation}
with
\be
g^{(0)}_{\rm lc}(\mathbf{x}_{\rm O},\mathbf{x}_{\rm S};\eta)=\frac{1}{4\pi}\frac{\rme^{-\ii\omega\Delta x}}{\Delta x},
\ee
and
\begin{align}
g^{(0)}_{\rm tail}(\mathbf{x}_{\rm O},\mathbf{x}_{\rm S};\eta)
=\frac{1}{4\pi}\int_0^{\Delta x}\ud\Delta\eta\,
B(\mathbf{x}_{\rm O},\eta,\mathbf{x}_{\rm S},\eta-\Delta\eta)\rme^{-\ii\omega\Delta\eta}.
\label{eq:G_frw_tail}
\end{align}

For LCDM cosmologies, the function $B$ is well approximated by \citep{Jokela:2022rhk}
\begin{align}
B(\eta_1,\eta_2)
={}&\frac{1}{\eta_1\eta_2}
+\frac{1}{(\eta_{\rm max}-\eta_1)(\eta_{\rm max}-\eta_2)}
-0.095H_0^2 \ ,
\end{align}
where the first term corresponds to the matter-dominated solution, the second to the de Sitter solution, and the constant term together with $\eta_{\rm max}=4.44H_0^{-1}$ was fixed by comparison with the numerical solution.

For the applications below it is useful to keep only the leading $H_0/\omega$ form of the tail. Starting from Eq.~\eqref{eq:G_frw_tail}, the tail integral may be integrated by parts:
\begin{align}
\int_0^{\Delta x}\ud\Delta\eta\,&B(\eta,\eta-\Delta\eta)\rme^{-\ii\omega\Delta\eta}
=\frac{B(\eta,\eta)}{\ii\omega}
-\frac{B(\eta,\eta-\Delta x)\rme^{-\ii\omega\Delta x}}{\ii\omega}\nonumber\\
&-\frac{1}{\ii\omega}\int_0^{\Delta x}\ud\Delta\eta\,\partial_{\eta_2}B(\eta,\eta_2)\big|_{\eta_2=\eta-\Delta\eta}\rme^{-\ii\omega\Delta\eta}.
\end{align}
Since $B\sim H_0^2$ and $\partial_{\eta_2}B\sim H_0^3$, the last term is suppressed by one additional power of $H_0/\omega$ relative to the endpoint terms.
Therefore, to leading order in $H_0/\omega$,
\begin{align}
g^{(0)}_\omega(\mathbf{x}_{\rm O},\mathbf{x}_{\rm S};\eta)
\simeq\frac{1}{4\pi}\left[
\frac{\rme^{-\ii\omega\Delta x}}{\Delta x}
+\frac{B(\eta,\eta)}{\ii\omega}
-\frac{B(\eta,\eta-\Delta x)\rme^{-\ii\omega\Delta x}}{\ii\omega}
\right] \ .
\end{align}

Using this Green's function, the leading correction in the potential is obtained from the Born approximation,
\begin{align}\label{eq:GreenOrderOneFRW}
\varphi_\omega&(\mathbf{x}_{\rm O},\mathbf{x}_{\rm S};\eta)
=\varphi^{(0)}_\omega(\mathbf{x}_{\rm O},\mathbf{x}_{\rm S};\eta)\nonumber\\
&+\omega^2\int \ud^3\mathbf{x}'\,
g^{(0)}_\omega(\mathbf{x}_{\rm O},\mathbf{x}';\eta)
V(\mathbf{x}';\eta)
\varphi^{(0)}_\omega(\mathbf{x}',\mathbf{x}_{\rm S};\eta) \ ,
\end{align}
where
\be
V(\mathbf{x}';\eta)\equiv-4U(\mathbf{x}')\left(1+\frac{a^2R^{\rm FLRW}}{6\omega^2}\right) \ .
\ee

Using Eq.~\eqref{eq:GreenFRW}, the leading light-cone contribution is
\begin{align}
\varphi^{(1)}_{\rm lc}
=\omega^2\int \ud^3\mathbf{x}'\,
g^{(0)}_{\rm lc}(\mathbf{x}_{\rm O},\mathbf{x}';\eta)
V(\mathbf{x}';\eta)
\varphi^{(0)}_{\rm lc}(\mathbf{x}',\mathbf{x}_{\rm S};\eta)\ ,
\end{align}
and the leading Hubble corrections arising from the single-tail cross terms are
\begin{eqnarray}
\varphi^{(1)}_{\rm tail,O}
=\omega^2\int \ud^3\mathbf{x}'\,
g^{(0)}_{\rm tail}(\mathbf{x}_{\rm O},\mathbf{x}';\eta)
V(\mathbf{x}';\eta)
\varphi^{(0)}_{\rm lc}(\mathbf{x}',\mathbf{x}_{\rm S};\eta)\ , \\ 
\varphi^{(1)}_{\rm tail,S}
=\omega^2\int \ud^3\mathbf{x}'\,
g^{(0)}_{\rm lc}(\mathbf{x}_{\rm O},\mathbf{x}';\eta)
V(\mathbf{x}';\eta)
\varphi^{(0)}_{\rm tail}(\mathbf{x}',\mathbf{x}_{\rm S};\eta)\ . \label{eq:tails}
\end{eqnarray}

The doubly-tail term is higher order in $H_0/\omega$ and will be neglected. In the limit $H_0/\omega\rightarrow0$, the tail contribution vanishes, and the flat-space functional form is recovered, as in the standard lensing approaches \citep{Bartelmann:1999yn}.

\subsection{Static black hole lens in FLRW}

We now consider how the static black-hole result is embedded in an expanding background. A convenient model is the McVittie spacetime, which describes a localized Schwarzschild-like mass in an FLRW universe. Although the full McVittie solution has limitations as a global model of an astrophysical black hole, it is sufficient for our purposes: we only use it in the weak-field region outside the lens, and only to leading order in the small expansion parameter $H/\omega$. In this regime, the metric captures the two ingredients needed here, namely the local black-hole potential and the cosmological redshifting of distances.

In the $H/\omega\rightarrow0$ limit, the results from Section~\ref{sec:Schw} can be translated straightforwardly to the McVittie metric, whose line element is
\begin{equation}
\begin{split}
ds^2={}&
-\left(
\frac{1-\frac{m}{2r}}{1+\frac{m}{2r}}
\right)^2dt^2
+a(t)^2
\left(1+\frac{m}{2r}\right)^4
\left(dr^2+r^2d\Omega^2\right),
\\
m(t)={}&\frac{GM}{a(t)} .
\end{split}
\end{equation}
One can remove the first derivative term of the equation of motion with the field redefinition $\phi_{\omega}= e^{m^2/(8 r^2)}\varphi/a(\eta)$, similar to the Schwarzschild case, to find
\begin{align}
&\left[\boldsymbol{\partial}^2+\omega^2\left( 1+a^2\frac{H'+2H^2}{\omega^2} \right) \right]\varphi
\simeq 4\omega^2 U \varphi \ ,
\nonumber\\
U=&
-\frac{m}{r}
\left(1+\ii \frac{a H}{\omega}+a^2\frac{H^2+H'}{\omega^2}\right) \nonumber\\
&-\frac{15m^2}{8r^2}
\left(
1+2\ii\frac{aH}{\omega}
+a^2\frac{H^2+31H'}{30\omega^2}
\right) -\frac{m^2}{16\omega^2r^4}
 \ ,
\label{eq:McVittie_eom}
\end{align}
where $H={a'(t)}/{a(t)}$, and $a=1$ at the observer's location. Thus, after neglecting the $H/\omega$ terms, the equation reduces to the Schwarzschild case with $G M \rightarrow m (t)$. In this limit, the FLRW result is obtained from the flat-space expressions by writing the three-dimensional vectors in terms of line-of-sight and transverse directions as in Eq.~\eqref{eq:3dto2d}. The transverse directions are unaffected by the expansion, while the line-of-sight separations are comoving distances. These are related to the angular diameter distances as $\chi_{\rm ls}=(1+z_{\rm S})D_{\rm ls}, \ \chi_{\rm so}=(1+z_{\rm S})D_{\rm so},$ and $\chi_{\rm lo}=(1+z_l)D_{\rm lo}$.  

Thus, the amplification factor in the small-angle limit is
\begin{equation}
\begin{split}
F\simeq{}&
1+\frac{r_{\rm E}^2}{|\Vec b|^2}
\exp\!\left(\ii\frac{|\Vec b|^2}{2r_{\rm F}^2}\right)
\Bigg[
1-\ii\frac{r_{\rm F}^2}{\chi_{\rm eff}^2}
\\
&+\frac{15\pi}{16}
\frac{|\Vec b|}{\chi_{\rm eff}}GM\omega
\left(
1-\ii\frac{r_{\rm F}^2}{2\chi_{\rm eff}^2}
-\frac{|\Vec b|^2}{240\chi_{\rm eff}^2}
\right)
\Bigg] .
\end{split}
\label{eq:FforFRW}
\end{equation}
where now write the effective distances as
\begin{equation}
r_{\rm F}=\sqrt{\frac{d_{\rm eff}}{\omega}}, \quad r_{\rm E}=\sqrt{4GMd_{\rm eff}}, \quad d_{\rm eff} =(1+z_l)\frac{D_{\rm lo}D_{\rm ls}}{D_{\rm so}} . \label{eq:scales}
\end{equation}
\begin{table*}
\centering
\small
\begin{tabular}{lccccccccccc}
\hline
Scenario & $f\,[{\rm Hz}]$ & $z_l$ & $z_{\rm S}$ & $M\,[M_\odot]$ & $r_{\rm E}\,[{\rm Mpc}]$ & $r_{\rm F}\,[{\rm Mpc}]$ & $r_{\rm E}^2/b^2$ & $b/r_{\rm F}$ & $\epsilon^{\rm ang}$ & $\nu $ & $\epsilon^{\rm H}$ \\
\hline
LIGO & $100$ & $0.2$ & $1$ & $100$ & $1.1\times10^{-7}$ & $9.9\times10^{-8}$ & $0.10$ & $3.5$ & $5.5\times10^{-10}$ & $0.31$ & $3.5\times10^{-21}$\\
L.H.T. & $100$ & $0.99$ & $1$ & $10^9$ & $6.9\times10^{-5}$ & $2.0\times10^{-8}$ & $0.10$ & $1.1\times10^4$ & $8.8\times10^{-6}$ & $3.1\times10^6$ & $3.5\times10^{-21}$\\
ET & $10$ & $0.5$ & $2$ & $10^3$ & $4.9\times10^{-7}$ & $4.4\times10^{-7}$ & $0.10$ & $3.5$ & $1.2\times10^{-9}$ & $0.31$ & $3.5\times10^{-20}$\\
LISA & $10^{-3}$ & $0.5$ & $2$ & $10^7$ & $4.9\times10^{-5}$ & $4.4\times10^{-5}$ & $0.10$ & $3.5$ & $1.2\times10^{-7}$ & $0.31$ & $3.5\times10^{-16}$\\
PTA & $10^{-8}$ & $0.5$ & $2$ & $10^9$ & $4.9\times10^{-4}$ & $0.014$ & $10^{-4}$ & $3.5$ & $3.9\times10^{-5}$ & $3.1\times10^{-4}$ & $3.5\times10^{-11}$\\
\hline
\end{tabular}
\caption{Illustrative lensing scenarios for several gravitational-wave bands. The L.H.T. row corresponds to a LIGO-band hierarchical triple where a binary of black holes orbits a super massive black hole. We note that the benchmark scenarios described here require the full Newtonian resummation from Eq.~\eqref{eq:exactNewt}, since the hierarchies in Eq.~\eqref{eq:Newt_scales} are not satisfied.}
\label{tab:gw_lensing_scales}
\end{table*}

\begin{table}
\centering
\small
\begin{tabular}{lccc}
\hline
Scenario & $\epsilon^{\rm GR}$ & $\epsilon_{\rm fd}$ & $\epsilon^{\rm Newt}$  \\
\hline
LIGO & $1.2\times10^{-10}$ & $2.4\times10^{-20}$ & $0.95$ \\
L.H.T. & $19$ & $6.3\times10^{-19}$ & $9.6\times10^{13}$ \\
ET & $2.7\times10^{-10}$ & $1.3\times10^{-19}$ & $0.95$ \\
LISA & $2.7\times10^{-8}$ & $1.3\times10^{-15}$ & $0.95$ \\
PTA & $8.6\times10^{-9}$ & $1.3\times10^{-10}$ & $9.5\times10^{-4}$ \\
\hline
\end{tabular}
\caption{Dimensionless control parameters for the five benchmark scenarios. All of these scenarios can be captured by the methods described here except the L.H.T. one. }
\label{tab:gw_lensing_dimensionless}
\end{table}

%--------------------------------
\subsubsection{Hubble corrections}
One could ask for the first corrections proportional to $H/\omega$. The contributions coming from the potential in Eq.~\eqref{eq:McVittie_eom} can be computed directly. At leading order, they give a correction to the amplification factor of the form
\begin{equation}
\Delta F^\text{FLRW}_{\mathcal{O}(G)} \sim \ii \frac{H_0}{\omega} \frac{ 2G M \chi_{\rm so}}{r_{\rm O} r_{\rm S} \bigl(1-\boldsymbol{\theta}_{\rm S}\!\cdot\!\boldsymbol{\theta}_{\rm O}\bigr)} e^{\ii\Omega^{\rm rf}}. \label{eq:ForderHomega}
\end{equation}
Even in the most optimistic PTA-band case, with frequencies of order nHz, we have $H_0/\omega\sim 10^{-10}$, so this is highly suppressed. We will analyze the order of magnitude of these quantities in more detail the next section.

The remaining Hubble corrections arise from the tail part of the FLRW Green's function. Since the zeroth-order Green's function and the unlensed waveform each split into a light-cone (flat space) and a tail piece, the Born correction contains a light-cone term, two single-tail cross terms, and a doubly-tail term. The doubly-tail term is higher order in $H/\omega$ and will be neglected. The leading tail terms are in Eq.~ \eqref{eq:tails}. We estimate these terms using the leading $H/\omega$ form of the tail Green's function derived above. For the observer-side contribution we find
\begin{align}
g^{(0)}_{\rm tail}(\mathbf{x}_{\rm O},\mathbf{x};\eta)
\simeq \frac{1}{4\pi \ii \omega}
\Big[
&B(\eta_{\rm O},\eta_{\rm O}) \nonumber \\
&-B(\eta_{\rm O},\eta_{\rm O}-|\mathbf{x}_{\rm O}-\mathbf{x}|)e^{-\ii\omega |\mathbf{x}_{\rm O}-\mathbf{x}|}
\Big] \ .
\end{align}
Taking $V(x)=4GM/r$ and expanding $|\mathbf{x}_{\rm S}-\mathbf{x}|\simeq r_{\rm S}+{\bf r}\cdot\boldsymbol{\theta}_{\rm S}$, $|\mathbf{x}_{\rm O}-\mathbf{x}|\simeq r_{\rm O}-{\bf r}\cdot\boldsymbol{\theta}_{\rm O}$, and $1/|\mathbf{x}_{\rm S}-\mathbf{x}|\simeq 1/r_S$, $1/|\mathbf{x}_{\rm O}-\mathbf{x}|\simeq 1/r_{\rm O}$, the result reduces to the same Fourier integrals used in the flat-space Born calculation. Keeping the leading terms gives
\begin{equation}
\phi^{(1)}_{\rm tail,O}\simeq-\ii\frac{GM}{\pi r_{\rm S}\omega}\left[\!B(\eta_{\rm O},\eta_{\rm O})e^{-\ii\omega r_{\rm S}}-\frac{B(\eta_{\rm O},\eta_{\rm O}-r_{\rm O})e^{-\ii\omega(r_{\rm O}+r_{\rm S})}}{2(1-\boldsymbol{\theta}_{\rm S}\!\cdot\!\boldsymbol{\theta}_{\rm O})}\right].
\end{equation}
The source-side term is obtained analogously:
\begin{equation}
\phi^{(1)}_{\rm tail,S}\simeq-\ii\frac{GM}{\pi r_{\rm O}\omega}\left[\!B(\eta_{\rm S},\eta_{\rm S})e^{-\ii\omega r_{\rm O}}-\frac{B(\eta_{\rm S},\eta_{\rm S}-r_{\rm S})e^{-\ii\omega(r_{\rm O}+r_{\rm S})}}{2(1-\boldsymbol{\theta}_{\rm S}\!\cdot\!\boldsymbol{\theta}_{\rm O})}\right].
\end{equation}
Since each endpoint coefficient is of order
\begin{equation}
B=H_0^2\times\mathcal{O}(1),
\end{equation}
the total single-tail contribution scales as
\begin{equation}
\phi^{(1)}_{\rm tail}\sim \ii \frac{GMH_0^2}{\omega r_{\rm SO}}\left(C_1+\frac{C_2}{1-\boldsymbol{\theta}_{\rm S}\!\cdot\!\boldsymbol{\theta}_{\rm O}}\right) \ ,
\end{equation}
where $C_i$ includes numerical factors irrelevant for our discussion, and the oscillating terms. This implies that the amplification factor in the small-angle limit will get a correction from the tail terms of order
\begin{equation}
\Delta F^\text{tail} \sim \ii \frac{H_0}{\omega}(H_0 \chi_{\rm eff}) \frac{r_{\rm E}^2}{|\Vec b|^2} \exp\!\left(\ii\frac{|\Vec b|^2}{2r_{\rm F}^2}\right) \ . \label{eq:ForderHomega2}
\end{equation}
While $H_0 \chi_{\rm eff}$ can be of order $0.1$ for cosmological lensing distances, we still have the large $H_0/\omega$ suppression.

%-----------------------
\subsection{Observational prospects and comparison between full three dimensional and paraxial approximation} \label{sec:observ}

\begin{table*}
\centering
\small
\begin{tabular}{lcccccccccccc}
\hline
Scenario & $f\,[{\rm Hz}]$ & $z_l$ & $z_{\rm S}$ & $M\,[M_\odot]$ & $r_{\rm E}\,[{\rm Mpc}]$ & $r_{\rm F}\,[{\rm Mpc}]$ & $b\,[{\rm Mpc}]$ & $b/r_{\rm F}$ & $\epsilon^{\rm ang}$ & $\nu$ & $\epsilon^{\rm H}$ \\
\hline
LISA-H.T. & $3.9\times10^{-4}$ & $0.5$ & $0.5$ & $1.0\times10^6$ & $1.5\times10^{-12}$ & $6.9\times10^{-12}$ & $3.6\times10^{-11}$ & $5.2$ & $3.0$ & $1.2\times10^{-2}$ & $9.0\times10^{-16}$\\
\hline
\end{tabular}
\caption{Hierarchical-triple benchmark in the LISA band. For this benchmark, the source is taken to be very close to the lens, so although we write $z_{\rm l}=z_{\rm S}=0.5$ in the table, this corresponds to a tiny separation of $\chi_{\rm ls}\simeq 1.2\times10^{-11}\,{\rm Mpc}\simeq 2.5\,{\rm AU}\simeq 42\,r_{\rm ISCO}$, where $r_{\rm ISCO}=6GM$ is the innermost stable circular orbit of the SMBH lens. Here $b$ is the lens-plane impact parameter, which we take to be $b\simeq 7.4\,{\rm AU}$.}
\label{tab:lisa_ht_scales}
\end{table*}

\begin{table}
\centering
\small
\begin{tabular}{lcccc}
\hline
Full 3d & $\epsilon^{\rm GR}$ & $\Omega^{\rm rf}$ & $\epsilon^{\rm fd}$ & $A$ \\
\hline
 & $9.9\times10^{-3}$ & $6.5$ & $0.11$ & $3.7\times10^{-3}$ \\
\hline
Naive 2d & $\frac{GM\omega\, b}{\sqrt{2}\,\chi_{\rm eff}}$ & $\frac{b^2}{2r_{\rm F}^2}$ & $\frac{r_{\rm F}^2}{\chi_{\rm eff}^2}$ & $\frac{r_{\rm E}^2}{b^2}$ \\
\hline
 & $2.5\times10^{-2}$ & $14$ & $0.33$ & $1.8\times10^{-3}$ \\
\hline
\end{tabular}
\caption{Dimensionless parameters controlling different expansions for the LISA hierarchical-triple benchmark. The upper block shows the full three-dimensional combinations appropriate for this configuration, while the lower block shows the corresponding naive small-angle quantities for comparison. Here, $A$ is the amplitude of the Born term contribution to the amplification factor in Eq.~\eqref{eq:FBorn}, which reduces to $r_{\rm E}^2/b^2$ in the small-angle limit.}
\label{tab:lisa_ht_dimensionless}
\end{table}

Having computed in the previous sections the amplification factor with the Born and post-Born corrections for a static compact lens, we now analyze the physical scenarios in which the approximations considered are valid, where they break down, and when the full three-dimensional analysis is relevant. To do so, we consider several benchmark scenarios for different gravitational wave detectors. Before analyzing those, we will clarify all the scales involved.

The meaning of each of the dimensionless parameters that play a role in the lensing dynamics can be summarized as follows:
\begin{center}
\fbox{%
\begin{minipage}{0.92\columnwidth}
\small
\begin{align*}
A&\equiv |F-1|
\sim \frac{r_{\rm E}^2}{b^2}
&&\text{Born amplitude}\\
\Omega^{\rm rf}
&\equiv \omega(r_{\rm O}+r_{\rm S}-\chi_{\rm so})
\sim \frac{b^2}{2r_{\rm F}^2}
&&\text{Fresnel phase}\\
\epsilon^{\rm fd}
&\equiv \frac{1}{\omega r_{\rm SO}}
\sim \frac{r_{\rm F}^2}{\chi_{\rm eff}^2}
&&\text{finite-distance correction}\\
\epsilon^{\rm GR}
&\equiv GM\omega\sqrt{1-\boldsymbol{\theta}_{\rm S}\!\cdot\!\boldsymbol{\theta}_{\rm O}}
\sim \frac{GM\omega|\Vec b|}{\sqrt{2}\chi_{\rm eff}}
&&\text{GR corrections}\\
\epsilon^{\rm Newt}
&\equiv\frac{(GM\omega)^2}{A} \sim \frac{(GM\omega)^2b^2}{r_{\rm E}^2} 
&&\text{Newtonian corrections}\\
\epsilon^H
&\equiv \frac{H_0}{\omega}
&&\text{FLRW corrections}\\
\epsilon^{\rm ang}
&\equiv \frac{|\Vec b|}{\chi_{\rm eff}}
&&\text{small-angle expansion} \\
\nu\equiv& GM\omega\lesssim 1  && \text{wave-optics regime}
\end{align*}
\end{minipage}}
\end{center}
where $\sim$ indicates the paraxial approximation.

For the illustrative examples in Table~\ref{tab:gw_lensing_scales}, we choose benchmark black-hole lens masses representative of different gravitational-wave bands. The angular-diameter distances entering the table are obtained from the corresponding lens and source redshifts assuming a flat $\Lambda$CDM cosmology with Planck-like parameters. The purpose of the table is not to provide an exhaustive survey, but rather to give a compact set of order-of-magnitude benchmarks illustrating how the relevant control parameters vary across frequency bands.

The LIGO hierarchical triple (L.H.T.) example is deep in the geometric-optics regime, with $GM\omega\gg1$. In this case the Fresnel phase is very large and the system is not expected to display the usual oscillatory features associated with wave optics. By contrast, the remaining examples satisfy $GM\omega\lesssim1$ and are in the wave-optics regime. As seen in Table~\ref{tab:gw_lensing_dimensionless}, the post-Born GR corrections are small in all scenarios except the L.H.T. example, and the Newtonian post-Born corrections are not suppressed in the LIGO, ET, LISA, and L.H.T. examples (but one can use the exact Newtonian solution as shown in Sec.~\ref{sec:Newt}).

Our computation is most interesting in the regime $\epsilon^{\rm ang}\gtrsim 1$, where the full three-dimensional geometry matters and one cannot reliably expand in the finite-distance phase. Finite-distance effects can become large when the lens is very close to the source, so that $\chi_{\rm eff}\sim\chi_{\rm ls}$ is small. This is precisely the situation that can arise in a hierarchical triple, where a compact binary orbits a much heavier compact object. While the LIGO band case in Table~\ref{tab:gw_lensing_scales} is beyond our approximations, a hierarchical triple in the LISA band, as shown in Table~\ref{tab:lisa_ht_scales}, can be described with our setup. Table~\ref{tab:lisa_ht_dimensionless} shows that this configuration lies in a particularly interesting intermediate regime: the Fresnel phase $\Omega^{\rm rf}$,
is order unity, the finite distance correction is not extremely small, $1/(\omega r_{\rm SO})\simeq 0.1$, and the GR correction parameter remains perturbative, $\epsilon^{\rm GR}\simeq 10^{-2}$. At the same time, the FLRW correction remains tiny, $H_0/\omega\ll 1$, so cosmological expansion effects are still strongly suppressed. Table~\ref{tab:lisa_ht_scales} makes clear that this regime is reached because the source--lens separation is very small, while also making the lens-plane impact parameter $b$ slightly larger than the effective comoving distance $\chi_{\rm eff}$.

The comparison with the naive small-angle quantities in Table~\ref{tab:lisa_ht_dimensionless} shows why the full three-dimensional geometry is important in this regime. The naive estimate would overestimate the Fresnel phase by roughly a factor of two. It would also underestimate the Born amplitude by about a factor of two, and similarly overestimate the finite-distance correction. The naive GR parameter is also larger by a factor of $\sim 2.5$. In this sense, hierarchical triples provide a natural configuration in which the genuinely three-dimensional effects captured by our calculation can become important. We note that with this setup we can go beyond the calculations of  \cite{Pijnenburg:2024btj} which assume an incoming plane wave, missing finite distance effects that are can become relevant, although in the present case we ignored polarization effects that can also become relevant.

\begin{figure}
  \centering
  \includegraphics[width=\columnwidth]{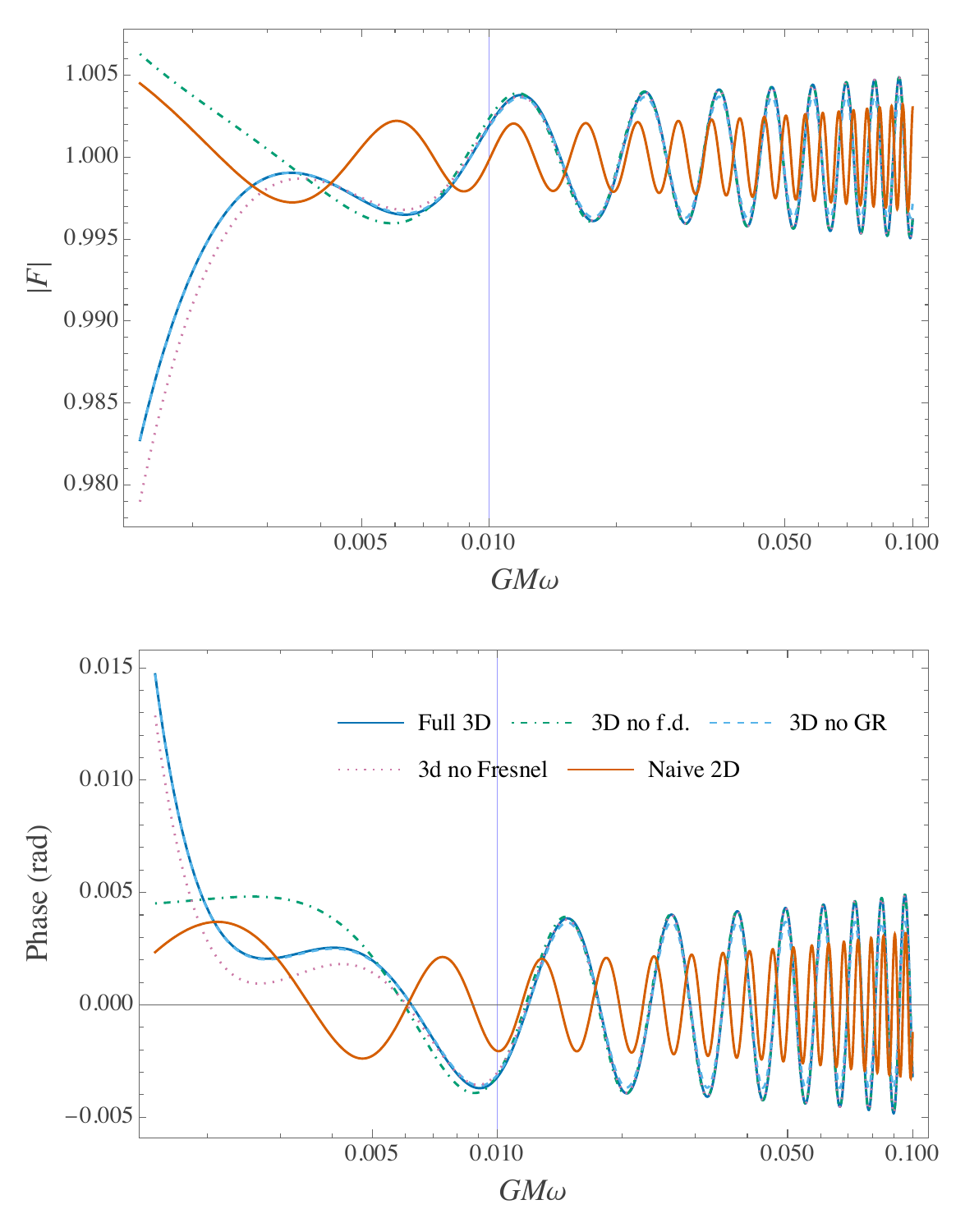}
 \caption{Comparison between the full three-dimensional amplification factor and the naive two-dimensional approximation for the controlled LISA hierarchical-triple benchmark with $M=10^6M_\odot$, $\chi_{\rm eff}=1.2\times10^{-11}\,{\rm Mpc}$ and $b=3.6\times10^{-11}\,{\rm Mpc}$. The vertical line marks the reference point $GM\omega=1.2\times10^{-2}$, corresponding to $f\simeq3.9\times10^{-4}\,{\rm Hz}$ which is used in Table~\ref{tab:lisa_ht_scales}. The upper panel shows $|F|$ and the lower panel shows the phase as functions of $GM\omega$. The full three dimensional curve is computed using the resummed finite-distance Born term together with the finite-distance Fresnel corrections through quartic order and the leading post-Born GR correction. The remaining curves show the effect of removing these contributions, and demonstrate that the naive 2d approximation fails in this regime.}
  \label{fig:lisa_ht_3d_vs_2d}
\end{figure}

The comparison in Fig.~\ref{fig:lisa_ht_3d_vs_2d} shows the different approximations analyzed in the paper. The solid blue curve uses the finite-distance-resummed Born and post-Born GR terms obtained using Eq.~\eqref{eq:Ip_shifted_resummation} and includes the Fresnel-phase corrections through $\Delta_4$ as defined in Appendix~\ref{app:fresnel}. The curve labeled ``3D no f.d.'' instead uses Eq.~\eqref{eq:FBorn} and Eq.~\eqref{eq:F_PB_GR} without the $1/(\omega r_{\rm SO})$ corrections and removes the genuinely finite-distance pieces in the Fresnel expansion, namely $\Delta_3$ and $\Delta_4$, while keeping the leading Fresnel corrections $\Delta_2$ and $\Delta_{22}$. The curve labeled ``3D no GR'' removes only the post-Born GR contribution, while the curve labeled ``3D no Fresnel'' keeps the finite-distance pieces but removes the leading Fresnel corrections $\Delta_2$ and $\Delta_{22}$. This separation is useful because the finite-distance corrections resummed by Eq.~\eqref{eq:Ip_shifted_resummation} are distinct from the Fresnel-phase expansion of Appendix~\ref{app:fresnel}: the former come from the Green's function prefactor, whereas the latter come from expanding the propagation phase beyond the leading term used in Eq.~\eqref{eq:FBorn}.

We observe that, at small $GM\omega$, the finite-distance expansion begins to break down, while the corrections to the Fresnel phase also become important. More significantly, the ``Naive 2D'' curve, which uses the paraxial approximation, is completely dephased relative to the full three-dimensional result, and its amplitude is also incorrect.

Finally, we emphasize that the statement that hierarchical triples may probe ``strong-field'' propagation, in the sense that the waves can pass within a few to tens of Schwarzschild radii of a massive companion, does not by itself determine the validity of the Born expansion used here. In the Green's function perturbative solution the relevant diagnostic is the size of the post-Born correction relative to the leading Born term. 

\section{Conclusions and future directions} \label{sec:concl}
We have shown that lensed waveforms can be computed systematically using Green's function methods and perturbative expansions. A central tool in the calculation is the reinterpretation of position-space integrals as Fourier transforms, or equivalently as momentum-space integrals whose solutions are well known from the scattering-amplitudes literature. This approach clarifies several approximations commonly used in the literature and provides a systematic way to go beyond them.

In particular, the method captures the full three-dimensional dynamics, including finite-distance corrections that become relevant when the lens is close to the source. We have produced new results for the three-dimensional GR corrections which agree in the small-angle limit with the literature. By carefully tracking the perturbative expansion, we have also expanded the result of \cite{Braga:2024pik} in the Born approximation to include subleading corrections in the Fresnel phase expansion and the finite distance corrections.  We also gave a closed-form resummation of the finite-distance $(\omega r_{\rm SO})^{-1}$ corrections arising from the Green's function prefactors (distinct from the thin-lens Fresnel resummation), allowing these effects to be included non-perturbatively, as used for the full three-dimensional curve in Fig. \ref{fig:lisa_ht_3d_vs_2d}. Additionally, we recovered the results from expanding the diffraction integral by resumming the Fresnel phase corrections in the paraxial approximation.

Finally, we highlighted that the finite distance and three-dimensional dynamics are especially relevant for hierarchical triples in the LISA band, as observed in Fig.~\ref{fig:lisa_ht_3d_vs_2d}. Recent work has shown that lensing in hierarchical triples can be polarization dependent, and that the wave-optics regime may require a treatment beyond a single scalar amplification factor \citep{Pijnenburg:2024btj}. Therefore, it would be interesting to apply the present method to the full spin-two case, where it could capture polarization-dependent and finite-distance corrections that are otherwise missed.

For the compact lens analyzed here, the curvature corrections associated with propagation in an FLRW Universe are always negligible. This need not remain true for smoothly distributed lenses, and our method gives a systematic way to include such effects in wave optics, including tail corrections.

It would also be interesting to extend the analysis to multiple lenses. Several studies of multi-lens wave optics use lens-plane or multi-plane diffraction-integral formalisms, and are therefore tied to the usual paraxial/small-angle setup \citep{Yamamoto:2003cd,Feldbrugge:2020tti,Ramesh:2021nnl}. In the Born approximation, the multiple-lens result is simply the linear superposition of the contribution from each lens, while the post-Born corrections are more interesting because they capture sequential lensing effects. These corrections should follow closely the calculation of Section~\ref{sec:Newt} for the Yukawa potential. In particular, the paraxial approximation is expected to break down when the lenses are sufficiently close to one another. Finally, it would be interesting to apply this method to other lens models, where it could help clarify the range of validity of commonly used approximations and provide a systematic way to go beyond them.

\section*{Acknowledgements}
We thank Alice Garoffolo and Pierre Fleury for useful discussions.  MCG is supported by the Imperial College Research Fellowship.

%%%%%%%%%%%%%%%%%%%% REFERENCES %%%%%%%%%%%%%%%%%%

\bibliographystyle{mnras}
\bibliography{refs}

\appendix

\section[Recovering the Born approximation from Braga et al.]{Recovering the Born approximation from a path integral approach} 
\label{app:BragaCalc}

We show how to recover the Born approximation from the path integral approach of \cite{Braga:2024pik}. In \cite{Braga:2024pik}, the frequency-domain Green's function is rewritten using a Schwinger proper-time representation. Instead of working directly with the waveform, one introduces an auxiliary evolution parameter $\tau$ and expresses the Green's function as
\begin{equation}
\phi_\omega({\bf x}_{\rm O},{\bf x}_{\rm S}) = -\frac{\ii}{\omega}\int_0^\infty d\tau\,e^{\ii\omega\tau}\, \tilde \phi_\omega({\bf x}_{\rm O},{\bf x}_{\rm S};\tau) \ .
\end{equation}
The potential is then treated perturbatively inside this kernel, and the usual Born expansion is recovered after performing the proper-time integrals.

The waveform function is expanded as
\begin{equation}
\phi_{\omega}({\bf x}_{\rm O},{\bf x}_{\rm S}) = \phi^{(0)}_{\omega}({\bf x}_{\rm O},{\bf x}_{\rm S})-\ii\omega \phi^{(1)}_{\omega}({\bf x}_{\rm O},{\bf x}_{\rm S}) + \ldots
\end{equation}
where the unlensed waveform, $\phi^{(0)}_{\omega}$ is given by Eq.~\eqref{eq:FreeProp} and the linear term is
\begin{equation}\label{eq:Propagator_factorization}
\phi^{(1)}_\omega({\bf x}_{\rm O},{\bf x}_{\rm S}) = -\frac{\ii}{\omega} \int_0^{+\infty}d\tau\, e^{\ii\omega \tau} \tilde \phi^{(1)}_\omega({\bf x}_{\rm O},{\bf x}_{\rm S}, \tau)
\end{equation}
with 
\begin{equation}
\label{eq:proper_time_second_order}
\begin{split}
\tilde \phi^{(1)}_\omega({\bf x}_{\rm O},{\bf x}_{\rm S}, \tau)= {}& \int_0^\tau d\tau_1\,\int_{-\infty}^{+\infty} d{\bf x} \,
\tilde G^{(0)}_\omega({\bf x}_{\rm O},{\bf x}, \tau- \tau_1 )\\
&\times 4U({\bf x})\tilde G^{(0)}_\omega({\bf x},{\bf x}_{\rm S}, \tau_1)
\end{split}
\end{equation}
and the proper time Green's function
\begin{equation}
\label{eq:free_particle_propagator_x_space} \tilde G_\omega^{(0)}({\bf x}_{\rm O},{\bf x}_{\rm S}, \tau )= \left( \frac{ \omega }{4 \ii \pi \tau} \right)^{3/2} \: e^{\frac{\ii \omega}{4\tau}|{\bf x}_{\rm O}-{\bf x}_{\rm S}|^2}\,.
\end{equation}
Hence the leading correction is 
\begin{align}
    \phi_\omega^{(1)}({\bf x}_{\rm O},{\bf x}_{\rm S} )&= -\frac{\ii}{\omega} \int_0^\infty d\tau\, e^{\ii\omega \tau} \int_0^\tau d\tau_1 \int d{\bf x}  \left(  \frac{ \omega  }{4 \ii \pi  (\tau-\tau_1)} \right)^{3/2} \: \nonumber\\
    & \times  e^{\frac{\ii \omega}{4(\tau-\tau_1)}|{\bf x}_{\rm O}-{\bf x}|^2}\,4U({\bf x})
    \left(  \frac{ \omega  }{4 \ii \pi  \tau_1} \right)^{3/2} \: e^{\frac{\ii \omega}{4\tau_1}|{\bf x}-{\bf x}_{\rm S}|^2}\,.
\end{align}
Using the integral \citep{Feynman:1965}
\begin{equation}
\begin{split}
&\int_0^T \exp\left(-\frac{a}{T-\tau} - \frac{b}{\tau}\right) \left[(T-\tau)\tau\right]^{-3/2} \, \mathrm{d}\tau \\
&\hspace{1cm}= \sqrt{\frac{\pi}{T^3}} \left( \frac{\sqrt{a}}{a} + \frac{\sqrt{b}}{b} \right)
\exp\left[ -\frac{1}{T} (\sqrt{a} + \sqrt{b})^2 \right] \, ,
\end{split}
\end{equation}
we can integrate over $\tau_1$ first, giving
\begin{align}
    \phi_\omega^{(1)}({\bf x}_{\rm O},{\bf x}_{\rm S} )&= \frac{\sqrt{i}\omega^{3/2}}{(4 \pi)^{5/2}} \int_0^\infty \frac{d\tau}{
    \tau^{3/2}}\, e^{\ii\omega \tau} \int d{\bf x} \,4U({\bf x})\, \nonumber\\
    &\mathrel{\phantom{=}}{}  \times \frac{(|{\bf x}_{\rm O}-{\bf x}|+|{\bf x}-{\bf x}_{\rm S}|)}{|{\bf x}_{\rm O}-{\bf x}|\,|{\bf x}-{\bf x}_{\rm S}|} e^{\frac{\ii \omega}{4\tau}(|{\bf x}_{\rm O}-{\bf x}|+|{\bf x}-{\bf x}_{\rm S}|)^2}.
\end{align}
The $\tau$ integral can similarly be performed using
\begin{equation}
\int_0^\infty \exp\left(-\frac{a}{\tau} - b\tau\right) \left(\tau\right)^{-3/2} \, \mathrm{d}\tau= \sqrt{\frac{\pi}{a}} \exp\left( -2\sqrt{a} \sqrt{b} \right) \,
\end{equation}
for $\text{Re}({b})>0$. One has to regularize the integral by taking $\omega\rightarrow \omega + \ii \epsilon$ and then take $\epsilon\rightarrow 0$ limit. Thus, we find
\begin{equation}
\begin{split}
\phi_\omega^{(1)}({\bf x}_{\rm O},{\bf x}_{\rm S} ) &=\frac{\ii \omega}{(4\pi)^2} \int\, d{\bf x}  \,\frac{4U({\bf x})}{|{\bf x}_{\rm O}-{\bf x}|\,|{\bf x}-{\bf x}_{\rm S}|} \\
&\mathrel{\phantom{=}}{}\times e^{\ii\omega(|{\bf x}_{\rm O}-{\bf x}|+|{\bf x}-{\bf x}_{\rm S}|)} \, .
\end{split}
    \label{eq:GreensFuncFirstOrder}
\end{equation}
This result applies to linear order to an arbitrary static potential which need not be localized in space. Hence, $-\ii\omega \phi^{(1)}_{\omega}({\bf x}_{\rm O},{\bf x}_{\rm S})$ corresponds to the Born correction.

\section{Fourier integrals} \label{app:fourier}
This appendix shows how to solve
\begin{align}
I^{(p)} &\equiv \int d^3\mathbf{r}\;
\frac{1}{|\mathbf{r}|^p}
\left( 1-\frac{\mathbf{r}\cdot\boldsymbol{\theta}_{\rm S}}{r_{\rm S}} +\frac{\mathbf{r}\cdot\boldsymbol{\theta}_{\rm O}}{r_{\rm O}}
\right) e^{\,\ii\mathbf{r}\cdot \mathbf{k}} \, ,
\end{align}
for $p$ a positive integer. Introducing
$\mathbf a \equiv \frac{\boldsymbol{\theta}_{\rm S}}{r_{\rm S}}-\frac{\boldsymbol{\theta}_{\rm O}}{r_{\rm O}}$,  we rewrite the integral as
\begin{align}
I^{(p)}=\left(1+\ii\,\mathbf a\!\cdot\!\nabla_{\mathbf k}\right)J_p(\mathbf k) \, , \quad \quad  J_p(\mathbf k)
\equiv \int d^3\mathbf r\;\frac{e^{\ii\mathbf r\cdot\mathbf k}}{|\mathbf r|^p} \, ,
\label{eq:Jp-def}
\end{align}
since $\ii\,\mathbf a\cdot\nabla_{\mathbf k}\,e^{\ii\mathbf r\cdot \mathbf k}=-\mathbf a\cdot\mathbf r\;e^{\ii\mathbf r\cdot \mathbf k}$. For $p\geq 3$ the integral \eqref{eq:Jp-def} is divergent at $r=0$, but it can be regularized by continuing to $d=3-2\epsilon$ dimensions. Doing so, one finds
\begin{align}
J_p^{\rm DR}(\mathbf k) &=
\mu^{-2\epsilon}\, 2^{3-p-2\epsilon}
\pi^{3/2-\epsilon} \frac{ \Gamma\!\left(\frac{3-p}{2}-\epsilon\right)}{\Gamma\!\left(\frac p2\right)}k^{p-3+2\epsilon} \ ,
\label{eq:Jp-master}
\end{align}
where $\mu$ is the standard dimensional-regularization scale introduced to keep dimensions fixed. For the finite cases $p=1,2$, one may set $\epsilon\to 0$ directly and obtain
\begin{align}
J_1(\mathbf k)=\frac{4\pi}{k^2}\ , \quad
J_2(\mathbf k)=\frac{2\pi^2}{k} \ .
\label{eq:J1J2}
\end{align}

For $p\geq 3$, the dimensionally regulated result \eqref{eq:Jp-master} should be understood as the analytic continuation of the Fourier transform. The structure depends on whether $p$ is even or odd. For $p$ even, $p=2m$ with integer $m\geq 2$, the limit $\epsilon\to 0$ is finite and one finds
\begin{align}
J_{2m}(\mathbf k)&= 2^{\,3-2m}\pi^{3/2} \frac{\Gamma\left(\frac32-m\right)}{\Gamma(m)}\,k^{\,2m-3}\ .
\label{eq:J-even}
\end{align}
The first few cases are
\begin{align}
J_4(\mathbf k)=-\pi^2 k\ ,
\quad
J_6(\mathbf k)=\frac{\pi^2}{12}k^3\ ,
\quad
J_8(\mathbf k)=-\frac{\pi^2}{360}k^5\ .
\label{eq:J-even-examples}
\end{align}
For $p$ odd, $p=2n+3$ with $n=0,1,2,\dots$, the gamma function develops a pole, since
\begin{align}
\Gamma(-n-\epsilon)= \frac{(-1)^{n+1}}{n!}
\left( \frac{1}{\epsilon}+\gamma_{\rm E}-H_n+O(\epsilon)
\right) \ ,
\label{eq:gamma-pole}
\end{align}
where $H_n=\sum_{m=1}^n \frac1m$ is the harmonic number. Expanding \eqref{eq:Jp-master} then gives
\begin{align}
J_{2n+3}^{\rm DR}(\mathbf k) &=C_n k^{2n} \Bigg[ \frac1\epsilon +\log\left(\frac{ k^2}{4\pi \mu^2}\right) +\gamma_{\rm E}-H_n +O(\epsilon) \Bigg], \nonumber\\
C_n&= \frac{(-1)^{n+1}\pi} {2^{\,n-1}n!(2n+1)!!}\ .
\label{eq:J-odd}
\end{align}
Equivalently, introducing the $\overline{\rm MS}$ scale $\bar\mu^2 \equiv 4\pi e^{-\gamma_{\rm E}}\mu^2$,
this may be written as
\begin{align}
J_{2n+3}^{\rm DR}(\mathbf k)=C_n k^{2n}
\left[\frac1\epsilon+\log\left( k^2 / \bar\mu^2\right)-H_n+O(\epsilon)\right]\ .
\label{eq:J-odd-msbar}
\end{align}
The first few odd cases are
\begin{align}
J_3^{\rm DR}(\mathbf k)&=-2\pi\left[\frac1\epsilon
+\log\!\left( k^2 / \bar\mu^2\right)\right]+O(\epsilon)\ ,
\label{eq:J3} \\
J_5^{\rm DR}(\mathbf k) &= \frac{\pi}{3}k^2 \left[ \frac1\epsilon +\log\!\left( k^2 / \bar\mu^2\right)-1
\right] +O(\epsilon)\ ,
\label{eq:J5} \\
J_7^{\rm DR}(\mathbf k) &= -\frac{\pi}{60}k^4 \left[ \frac1\epsilon +\log\!\left( k^2 / \bar\mu^2\right)-\frac32
\right] +O(\epsilon)\ .
\label{eq:J7}
\end{align}
The pole $1/\epsilon$ signals the short-distance ultraviolet divergence of the Fourier integral for $p$ odd with $p\geq 3$. In the $\overline{\rm MS}$ renormalization scheme one subtracts the $\epsilon$ pole in the formulas above.

We next compute the derivative part. Since $J_p(\mathbf k)$ depends only on $k=|\mathbf k|$, rotational invariance implies
\begin{align}
\nabla_{\mathbf k}J_p(k) = \hat{\mathbf k}\,\frac{dJ_p(k)}{dk}, \qquad \hat{\mathbf k}\equiv \frac{\mathbf k}{k}\ .
\label{eq:radial-grad}
\end{align}
Equation \eqref{eq:Jp-def} therefore becomes
\begin{align}
I^{(p)} = J_p(k) + \ii\,(\mathbf a\!\cdot\!\hat{\mathbf k})\,\frac{dJ_p(k)}{dk} \ .
\label{eq:Ip-general}
\end{align}
Whenever $J_p(k)\propto k^\alpha$, this simplifies to
\begin{align}
I^{(p)} = J_p(k)\left(1+\ii\,\alpha\,\frac{\mathbf a\!\cdot\!\hat{\mathbf k}}{k}\right)\ .
\label{eq:Ip-powerlaw}
\end{align}

In particular, for the convergent cases \eqref{eq:J1J2} one obtains
\begin{align}
I^{(1)} &= \frac{4\pi}{k^2} \left( 1-\frac{2\ii\,\mathbf a\!\cdot\!\hat{\mathbf k}}{k} \right), \label{eq:I1} \\
I^{(2)} &= \frac{2\pi^2}{k} \left( 1-\frac{\ii\,\mathbf a\!\cdot\!\hat{\mathbf k}}{k} \right) \ . \label{eq:I2}
\end{align}
For $p$ even with $p=2m\geq 4$, using \eqref{eq:J-even} one similarly finds
\begin{align}
I^{(2m)} = J_{2m}(k) \left[ 1+\ii\,(2m-3)\frac{\mathbf a\!\cdot\!\hat{\mathbf k}}{k} \right] \ . \label{eq:I-even}
\end{align}
Meanwhile, for $p$ odd, $p=2n+3$,
\begin{align}
\frac{\ud J_{2n+3}^{\rm DR}(k)}{\ud k}
&=C_n k^{2n-1}\left\{2n\left[\frac1\epsilon+\log\!\left( k^2 / \bar\mu^2\right)-H_n\right]+2\right\}
\nonumber\\
&\mathrel{\phantom{=}}{}+O(\epsilon)\ .
\label{eq:dJodd}
\end{align}
and therefore
\begin{align}
I^{(2n+3)}&=J_{2n+3}^{\rm DR}(k)
+\ii(\mathbf a\!\cdot\!\hat{\mathbf k})\,C_n k^{2n-1}\nonumber\\
&\mathrel{\phantom{=}}{}\times\left\{2n\left[\frac1\epsilon+\log\!\left( k^2 / \bar\mu^2\right)-H_n\right]+2\right\}+O(\epsilon)\ .
\label{eq:I-odd}
\end{align}

After performing the $\overline{\rm MS}$ subtraction, one drops the $1/\epsilon$ pole and obtains
\begin{align}
I^{(2n+3)}_{\overline{\rm MS}}&= C_n k^{2n} \Bigg[
\log\!\left( k^2 / \bar\mu^2\right) -H_n \nonumber \\ 
&\mathrel{\phantom{=}}{}+\ii\frac{\mathbf a\!\cdot\!\hat{\mathbf k}}{k} \left\{ 2n \left[ \log\!\left( k^2 / \bar\mu^2\right) -H_n
\right] +2 \right\} \Bigg] \ .
\end{align}
The appearance of the logarithm reflects the short-distance sensitivity of the Born integral for sufficiently singular terms in the near-lens expansion. In dimensional regularization, the pole and the associated scale dependence are absorbed into local finite-size counterterms for the compact object \citep{Goldberger:2004jt}. Equivalently, these terms do not represent new long-distance propagation effects: their scheme dependence is canceled by short-distance coefficients that encode the structure of the lens at radii of order the Schwarzschild scale. For the applications considered here, for which the wave probes impact parameters much larger than this scale ($G M\ll b$), these renormalized finite-size contributions are not resolved and can be ignored for our purposes.

For the cases of interest in the main text we have $ \mathbf{k}=\omega(\boldsymbol{\theta}_{\rm S}-\boldsymbol{\theta}_{\rm O})$, which implies
\begin{align}
\frac{\mathbf a\!\cdot\!\hat{\mathbf k}}{k} = \frac{1}{2\omega}\left(\frac1{r_{\rm S}}+\frac1{r_{\rm O}}\right)\equiv\frac{1}{2\omega r_{\rm SO}} \ .
\end{align}

\section{Finite-distance resummation}
\label{app:finite_distance_resummation}

We summarize the formal resummation of the finite-distance corrections that appear in the Born integral. The relevant integral is
\begin{equation}
I^{(p)} \equiv r_{\rm O} r_{\rm S} \int \ud^3\mathbf r\, \frac{e^{\ii\mathbf k\cdot\mathbf r}}{|\mathbf r|^p} \frac{1}{|\mathbf x_{\rm O}-\mathbf r|\,|\mathbf r-\mathbf x_{\rm S}|} .
\end{equation}
In the finite-distance expansion used in the main text, we write
\begin{eqnarray}
\frac{1}{|\mathbf x_{\rm O}-\mathbf r|\,|\mathbf r-\mathbf x_{\rm S}|}
\simeq
\frac{1}{r_{\rm O} r_{\rm S}}
\frac{1}{1+\mathbf r\cdot \frac{\boldsymbol\theta_{\rm S}}{r_{\rm S}}}
\frac{1}{1-\mathbf r\cdot \frac{\boldsymbol\theta_{\rm O}}{r_{\rm O}}} \ ,
\end{eqnarray}
Using Eq.~\eqref{eq:Jp-def} and following the same procedure as in Appendix~\ref{app:fourier}, this can be written as the formal operator identity
\begin{equation}
I^{(p)} = \frac{1}{1-\ii \ \frac{\boldsymbol\theta_{\rm S}}{r_{\rm S}}\cdot\nabla_{\mathbf k}} \frac{1}{1+\ii \ \frac{\boldsymbol\theta_{\rm O}}{r_{\rm O}}\cdot\nabla_{\mathbf k}} J_p(k) . \label{eq:Ip_operator_resummation}
\end{equation}
Equivalently, using
\begin{eqnarray}
\frac{1}{1-\ii \frac{\boldsymbol\theta_{\rm S}}{r_{\rm S}}\cdot\nabla_{\mathbf k}}
=
\int_0^\infty \ud\alpha\,e^{-\alpha}e^{\ii\alpha \frac{\boldsymbol\theta_{\rm S}}{r_{\rm S}}\cdot\nabla_{\mathbf k}},
\\
\frac{1}{1+\ii \ \frac{\boldsymbol\theta_{\rm O}}{r_{\rm O}}\cdot\nabla_{\mathbf k}}
=
\int_0^\infty \ud\beta\,e^{-\beta}e^{-\ii \beta \frac{\boldsymbol\theta_{\rm O}}{r_{\rm O}}\cdot\nabla_{\mathbf k}},
\end{eqnarray}
and considering $J_p(k)=C_p k^s$, we obtain
\begin{equation}
I^{(p)} = C_p \int_0^\infty \ud\alpha \int_0^\infty \ud\beta\, e^{-\alpha-\beta} \left| \mathbf k+\ii\alpha \frac{\boldsymbol\theta_{\rm S}}{r_{\rm S}}-\ii\beta \frac{\boldsymbol\theta_{\rm O}}{r_{\rm O}} \right|^s \ . \label{eq:Ip_shifted_resummation}
\end{equation}
We can now substitute the kinematics used in the main text,
$\mathbf k=\omega(\boldsymbol\theta_{\rm S}-\boldsymbol\theta_{\rm O})$, together with the definition $c\equiv \boldsymbol\theta_{\rm S}\cdot\boldsymbol\theta_{\rm O}$ to obtain
\begin{align}
\left|\mathbf{k}+\ii\alpha \frac{\boldsymbol\theta_{\rm S}}{r_{\rm S}}-\ii\beta \frac{\boldsymbol\theta_{\rm O}}{r_{\rm O}} \right|^2 &= 2\omega^2 (1-c)\Bigg[ 1 + \frac{\ii}{\omega}
\left( \frac{\alpha}{r_{\rm S}} + \frac{\beta}{r_{\rm O}} \right) \nonumber \\
&\mathrel{\phantom{=}}{}-\frac{1}{2\omega^2(1-c)}\left( \frac{\alpha^2}{r_{\rm S}^2} + \frac{\beta^2}{r_{\rm O}^2} - 2c\frac{\alpha\beta}{r_{\rm S} r_{\rm O}}
\right) \Bigg].
\end{align}
Expanding the resummed expression to first order gives
\begin{equation}
I^{(p)} = J_p(k) \left[ 1+\frac{\ii s}{2\omega r_{\rm SO}} +\cdots \right] \ ,
\end{equation}
which reproduces the result from Appendix~\ref{app:fourier}. 

In the small-angle limit, one can further simplify this to
\begin{equation}
\left|\mathbf{k}+\ii\alpha \frac{\boldsymbol\theta_{\rm S}}{r_{\rm S}}-\ii\beta \frac{\boldsymbol\theta_{\rm O}}{r_{\rm O}} \right|^2 \simeq \frac{\omega^2 |\Vec{b}|^2}{\chi_{\rm eff}^2} +\ii\omega\frac{|\Vec{b}|^2}{\chi_{\rm eff}^2} \left( \frac{\alpha}{\chi_{\rm ls}} +\frac{\beta}{\chi_{\rm lo}} \right) -\left( \frac{\alpha}{\chi_{\rm ls}} -\frac{\beta}{\chi_{\rm lo}} \right)^2 .
\end{equation}

This resummation should not be confused with the thin-lens Fresnel resummation leading to the exponential-integral result of \cite{Takahashi:2005sxa}, which follows from the Born approximation of the exact Newtonian result \citep{Peters:1974gj}. Here we have only resummed the finite-distance expansion of the Green's function prefactor,
$|\mathbf{x}_{\rm O}-\mathbf r|^{-1}|\mathbf r-\mathbf{x}_{\rm S}|^{-1}$, while the usual thin-lens result keeps the quadratic transverse phase unexpanded and projects the three-dimensional potential onto the lens plane. That resummation is derived in Appendix~\ref{app:fresnel}.

\section{Fresnel phase resummation} \label{app:fresnel}

We revisit the finite-distance corrections to the Fresnel phase in the Born integral. In Sec.~\ref{sec:Fresnel} we kept the first correction from the quadratic term in the propagation phase in Eq.~\eqref{eq:phase_exp}. Keeping the same expansion to quartic order gives
\begin{equation*}
\widehat{I}^{(1)} \simeq \frac{2\pi}{\omega^2(1-\boldsymbol\theta_{\rm S}\cdot\boldsymbol\theta_{\rm O})} \left( 1+\Delta_2+\Delta_3+\Delta_{22}+\Delta_4+\cdots \right) \ ,
\end{equation*}
where $\Delta_{22}$ is the term generated by expanding the quadratic phase to second order. Explicitly,
\begin{equation}
\Delta_2 = -\frac{\ii}{\omega r_{\rm SO}}\, \frac{\boldsymbol\theta_{\rm S}\cdot\boldsymbol\theta_{\rm O}}{ 1-\boldsymbol\theta_{\rm S}\cdot\boldsymbol\theta_{\rm O}} \ ,
\end{equation}
\begin{equation}
\Delta_3 = -\frac{1}{\omega^2} \left(\frac{1}{r_{\rm O}^2}+\frac{1}{r_{\rm S}^2}\right) \frac{3\boldsymbol\theta_{\rm S}\cdot\boldsymbol\theta_{\rm O}+1}{ 1-\boldsymbol\theta_{\rm S}\cdot\boldsymbol\theta_{\rm O}}\ ,
\end{equation}
\begin{equation}
\begin{split}
\Delta_{22}
=
-&\frac{1}{
\omega^2(1-\boldsymbol\theta_{\rm S}\cdot\boldsymbol\theta_{\rm O})^2}
\bigg(
\frac{3(\boldsymbol\theta_{\rm S}\cdot\boldsymbol\theta_{\rm O})^2-1}{r_{\rm O}^2}
\\
&+
\frac{(\boldsymbol\theta_{\rm S}\cdot\boldsymbol\theta_{\rm O})^2+3}{r_{\rm O}r_{\rm S}}
+
\frac{3(\boldsymbol\theta_{\rm S}\cdot\boldsymbol\theta_{\rm O})^2-1}{r_{\rm S}^2}
\bigg)\ ,
\end{split}
\end{equation}
and
\begin{equation}
\Delta_4 = -\frac{3\ii}{\omega^3} \left(\frac{1}{r_{\rm O}^3}+\frac{1}{r_{\rm S}^3}\right) \frac{5(\boldsymbol\theta_{\rm S}\cdot\boldsymbol\theta_{\rm O})^2 -\boldsymbol\theta_{\rm S}\cdot\boldsymbol\theta_{\rm O}-2 }{ (1-\boldsymbol\theta_{\rm S}\cdot\boldsymbol\theta_{\rm O})^2} \ .
\end{equation}
We now take the small-angle limit. Using
\begin{equation}
1-\boldsymbol\theta_{\rm S}\cdot\boldsymbol\theta_{\rm O} \simeq \frac{b^2}{2\chi_{\rm eff}^2} \ , \qquad r_{\rm SO}\simeq \chi_{\rm eff}\ ,
\end{equation}
we find
\begin{equation}
\Delta_2 \simeq -2\ii\,\frac{r_{\rm F}^2}{b^2}\ , \qquad \Delta_{22} \simeq -8\,\frac{r_{\rm F}^4}{b^4}\ ,
\end{equation}
while
\begin{equation}
\Delta_3 \simeq -8\,\frac{r_{\rm F}^4}{b^2} \left(\frac{1}{\chi_{\rm lo}^2}+\frac{1}{\chi_{\rm ls}^2}\right)\ , \qquad \Delta_4 \simeq -24\ii\,\frac{r_{\rm F}^6\chi_{\rm eff}}{b^4} \left(\frac{1}{\chi_{\rm lo}^3}+\frac{1}{\chi_{\rm ls}^3}\right)\ .
\end{equation}
Therefore, the leading small-angle terms are
\begin{equation}
\widehat{I}^{(1)}  \simeq \frac{4\pi\chi_{\rm eff}^2}{\omega^2b^2} \left( 1 -2\ii\,\frac{r_{\rm F}^2}{b^2} -8\,\frac{r_{\rm F}^4}{b^4} +\cdots \right)\ .
\end{equation}
The term $\Delta_3$ has not been included in the last expression because it is suppressed relative to $\Delta_{22}$ by powers of $b^2/\chi^2$. This shows the general small-angle hierarchy: linear higher-order terms in the phase carry additional inverse powers of the line-of-sight distances, while repeated insertions of the quadratic phase generate the powers of $r_{\rm F}^2/b^2$.

We can resum the contribution from the quadratic phase in the small-angle limit. Keeping only the quadratic phase gives
\begin{equation}
\widehat{I}^{(1)} \simeq \exp\left[ -\frac{\ii \omega}{2} Q_{ij}\frac{\partial^2}{\partial k_i\partial k_j} \right] \frac{4\pi}{k^2}.
\end{equation}
In the small angle limit, using $\mathbf{k}=\omega(\boldsymbol{\theta}_{\rm S}-\boldsymbol{\theta}_{\rm O})$, only the transverse lens-plane components contribute at leading order. We denote these two-dimensional transverse indices by $a,b=1,2$. Then we have
\begin{equation}
Q_{ab}\simeq \frac{1}{\chi_{\rm eff}}\delta_{ab}, \quad k_a\simeq -\frac{\omega}{\chi_{\rm eff}}b_a , \ \rightarrow \ \frac{\partial}{\partial k_a} = -\frac{\chi_{\rm eff}}{\omega}\frac{\partial}{\partial b_a} \ .
\end{equation}
Hence, the quadratic operator reduces to
\begin{equation}
-\frac{\ii \omega}{2} Q_{ij}\frac{\partial^2}{\partial k_i\partial k_j} \simeq -\frac{\ii}{2}r_{\rm F}^2\,\partial_{\bf b}^2 .
\end{equation}
Thus
\begin{equation}
\widehat{I}^{(1)} \simeq \frac{4\pi\chi_{\rm eff}^2}{\omega^2} {\rm e}^{ -\frac{\ii}{2}r_{\rm F}^2\partial_{\bf b}^2} \frac{1}{b^2}\simeq \frac{4\pi\chi_{\rm eff}^2}{\omega^2b^2} \sum_{n=0}^{\infty} n!\left(-2\ii\,\frac{r_{\rm F}^2}{b^2}\right)^n \ ,
\end{equation}
where we used that away from $b=0$,
\begin{equation}
(\partial_{\bf b}^2)^n\frac{1}{b^2} = 4^n(n!)^2\frac{1}{b^{2n+2}} \ .
\end{equation}
The series is asymptotic and can be identified with the principal branch of the exponential integral as
\begin{equation}
\sum_{n=0}^{\infty}n!\left(-2\ii\,\frac{r_{\rm F}^2}{b^2}\right)^n \sim -\ii\,\frac{b^2}{2r_{\rm F}^2} e^{-\ii b^2/(2r_{\rm F}^2)} E_1\!\left(-\ii\,\frac{b^2}{2r_{\rm F}^2}\right),
\end{equation}
where $E_1$ is taken on its principal branch. Therefore
\begin{equation}
F\simeq 1-2\ii G M \omega E_1\left(-\ii\frac{b^2}{2r_{\rm F}^2}\right) \ ,
\end{equation}
which agrees with \cite{Takahashi:2005sxa}. We note that we were not able to find a similar resummation for the full three-dimensional case.

\label{lastpage}
\end{document}